\documentclass[aps,pra,twocolumn,floats,showpacs,superscriptaddress]{revtex4-2}
\usepackage{graphicx,amsmath,amsfonts,amssymb,upgreek,txfonts, color}
\usepackage[colorlinks,linkcolor=blue,citecolor=blue,urlcolor=blue,breaklinks=true]{hyperref}
\usepackage{color, colortbl}
\definecolor{lightgreen}{rgb}{0.88,1,1}
\begin{document}
\title{Enhanced Sensing by Geometric Tuning of YIG Spheres: Noise Reduction, Signal Amplification and Directional Magnetic Field Detection}
\author{Zheng Liu, Ding-hui Xu, Yi-jia Yang, and Chang-shui Yu}
\email{Electronic address: ycs@dlut.edu.cn}
\address{School of Physics, Dalian University of Technology, Dalian 116024,
	P.R. China}
\begin{abstract}
	Noise suppression and directional signal enhancement are essential challenges in detecting weak magnetic fields in cavity electrodynamics systems. Traditional schemes struggle to reduce magnonic probe noise but lack directional sensing capabilities. We exploit an innovative and intrinsic squeezing mechanism by leveraging the geometric configuration of an anisotropic ellipsoidal yttrium iron garnet (YIG) sphere and its interaction with internal demagnetization fields. This mechanism can enhance magnetic field signals and suppress noise in the target direction while suppressing sensitivity in non-target directions to avoid disturbing the target direction,  thus generating a directionally selective sensing scheme realizing high-precision detection in complex environments.  In particular, the target-direction sensor performance can be optimized by adjusting the YIG sphere's geometry (e.g., aspect ratio)  without complex setups, ensuring high feasibility and scalability. Our approach offers greater flexibility and directionality by tuning the YIG sphere's geometry than existing methods. This innovation provides a new approach for weak magnetic field detection in cavity magnonics systems, with potential applications in biomedical imaging, quantum sensing, precision measurement, and environmental monitoring.
	\end{abstract}
	\maketitle
\section{Introduction}
	\label{I} 
In recent years, quantum techniques for enhancing the sensitivity of physical quantity sensing have gained widespread attention. Leveraging advanced technologies such as squeezing, entanglement, coherence, and indefinite causal order, quantum sensing has demonstrated remarkable potential to achieve high sensitivity in detecting weak signals, thereby significantly improving sensor performance \cite{RevModPhys.89.035002, Yin2023, PhysRevA.110.033505, PhysRevLett.108.120801, PhysRevLett.129.070502, PhysRevLett.127.113601,https://doi.org/10.1002/andp.202100421, Li:18,https://doi.org/10.1002/adma.201401144,PhysRevApplied.19.044044,PhysRevB.99.214415,PhysRevA.103.062605,PhysRevA.106.013506,Schliesser_2008,PhysRevA.82.061804,10.1063/1.5055029,PhysRevLett.125.147201}.
Among the various branches of quantum sensing, quantum magnetometry has emerged as a prominent focus due to its critical applications in fields such as biological sciences \cite{s20061569}, geophysics \cite{Stolz_2021}, and dark matter detection \cite{PhysRevD.99.075031}. Numerous magnetometers, including optically pumped atomic magnetometers \cite{PhysRevLett.130.023201}, NV center magnetometers \cite{PhysRevApplied.11.034029}, and superconducting quantum interference devices (SQUIDs) \cite{10.1063/5.0065790},  have already demonstrated quantum-enhanced sensitivity.
In particular, cavity optomechanical magnetometers and cavity magnonic magnetometers are attracting growing research interest due to their significant advantages in miniaturization and integration \cite{Li:18, https://doi.org/10.1002/adma.201401144,PhysRevA.103.062605,PhysRevB.99.214415}.

Recent breakthroughs in ferrimagnetic and antiferromagnetic materials, particularly yttrium iron garnet (YIG) spheres, have garnered substantial attention in quantum information processing. These materials are distinguished by their exceptional spin density \cite{CHEREPANOV199381}, prolonged decoherence times \cite{PhysRevApplied.2.054002, PhysRevB.93.144420}, and excellent tunability in the high-frequency range \cite{PhysRevB.96.094412}, offering significant advantages over conventional hybrid systems.
In YIG spheres, the fundamental spin wave mode, known as the Kittel mode, represents a quantized magnon mode that can strongly interact with various fields, such as optical and microwave fields, making it an ideal candidate for quantum sensing readout \cite{PhysRevLett.104.077202, PhysRevLett.117.133602, PhysRevLett.116.223601, PhysRevLett.117.123605, PhysRevA.94.033821,PhysRevA.105.033507,PhysRevA.103.062605}. Moreover, coupling magnons with superconducting qubits significantly enhances the precision of spin readout \cite{PhysRevA.106.053714, PhysRevLett.125.117701, PhysRevA.101.042331}.
These advancements have enabled various innovative applications, including the generation of non-classical magnon states \cite{PhysRevLett.121.203601, PhysRevResearch.1.023021, PhysRevLett.124.053602, PhysRevLett.127.087203,PhysRevA.110.053710}, nonreciprocal signal transport \cite{PhysRevApplied.12.034001, PhysRevLett.123.127202, PhysRevA.101.043842, PhysRevLett.124.107202, PhysRevA.105.013711}, quantum networks \cite{PhysRevA.108.043703}, and contributions to the exploration of non-Hermitian quantum physics \cite{PhysRevLett.121.137203, PhysRevB.100.094415, PhysRevLett.125.147202, PhysRevA.103.063708}. These developments promise significant improvements in measurement precision across various scientific fields and lay an important foundation for the advancement of quantum technologies.
	
However,  the sensing performance of cavity magnonic systems is critically limited by quantum noise and probe (magnon) input thermal noise. These noise sources can overwhelm the signal, making high-precision sensing tasks unfeasible \cite{RevModPhys.82.1155, Safavi-Naeini_2013, zhang}.  Quantum noise in cavity-magnonic systems includes shot noise from photons within the microwave cavity and backaction noise arising from photon-magnon coupling interactions, which define the standard quantum limit (SQL) \cite{RevModPhys.86.1391, PhysRevA.103.062605}. To surpass the SQL, various strategies have been proposed to reduce quantum noise from the cavity field, such as coherent quantum noise cancellation \cite{PhysRevLett.105.123601, PhysRevX.2.031016, PhysRevA.92.043817, Motazedifard_2016, vitali, PhysRevA.89.053836}, quadratic coupling \cite{PhysRevA.102.063523, Chao:21, Zhang:24}, and quantum squeezing \cite{zhengjiaoguanlian, Zhao2019, Zhang2024,10.1063/5.0208107, PhysRevA.109.023709}, etc.,  but enhancing directional signals or suppressing probe input thermal noises remain a challenge, needless to say, their integration.

In this paper, we propose a novel scheme to reduce cavity field quantum noise and magnon mode input thermal noise, enhancing the performance of cavity magnonic weak magnetic field sensing by leveraging the anisotropy of the elliptical YIG sphere. The anisotropic squeezing induced by the spatial distribution of the elliptical YIG sphere not only effectively suppresses probe input thermal noise and cavity field quantum noise but also enhances the system's response to input signals. More importantly, the spatially dependent squeezing effect introduces directional dependence, enabling directional magnetic field sensing. This allows for selective detection of magnetic fields from specific directions while suppressing disturbance from signals in other directions. The noise from the target direction is redirected to other directions through this squeezing effect, providing a significant advantage for applications such as directional sensing and noise-resistant magnetic field detection. Additionally, we investigate the impact of other key parameters on weak magnetic field sensing performance and elucidate the physical mechanisms behind the suppression of cavity quantum noise and magnon input thermal noise within the sensing system.

The structure of this paper is organized as follows.  Section \ref{sec2}  presents the proposed sensing scheme, encompassing the system model and Hamiltonian.  Then, the dynamics and the phase quadrature output are given for the weak magnetic field sensing system. In Section \ref{sec3}, the quantum noise spectrum of the system output is evaluated using homodyne detection. The sensing performance is assessed regarding the system's response to external signals, cavity field quantum noise, and magnon input  thermal noise.The underlying physical mechanisms, as well as the signal-to-noise ratio (SNR) and sensitivity, are further detailed in Sections \ref{sec4} and \ref{sec5}. Finally, Section \ref{sec6} summarizes the paper with conclusions and discussions about its broader implications. More detailed information is provided in the appendix section.
	\section{\label{sec2}Weak Magnetic Field Sensing Model, Hamiltonian, and System Dynamics}
	\begin{figure}
	\centering\includegraphics[width=9.5cm,height=4cm]{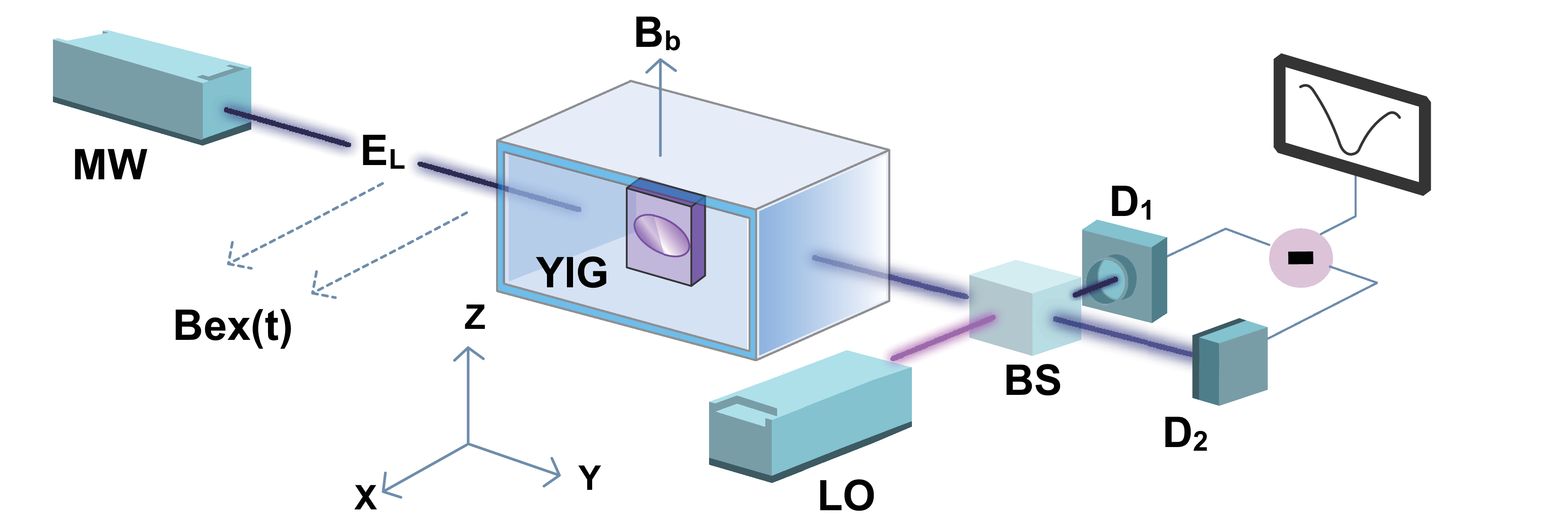}
		\caption{A schematic diagram of weak magnetic field sensing in an anisotropic cavity magnonic system. An anisotropic elliptical YIG (yttrium iron garnet) sphere serves as the probe for detecting external magnetic field signals with its long axis oriented along the x-axis. The microwave signal output from the cavity passes through a microwave beam splitter (BS) and is subsequently detected by two microwave photon detectors ($D_1$ and $D_2$). The signal is then processed via modulation with a local oscillator (LO) and undergoes spectral analysis of the cavity field's quadrature components. The phase angle of the LO is set to $\pi/2$ to probe the phase quadrsture component in the output spectrum of the cavity field. The external bias magnetic field, $\rm B_b$, is oriented along the z-axis, while the measured magnetic field, $\rm B_{ex}(t)$, is aligned along the x-axis. The amplitude of the externally driven microwave field in the cavity is denoted as $\rm E_L$.}
		\label{Fig1}
	\end{figure}	
As depicted in Fig. \ref{Fig1}, the weak magnetic field sensing system comprises an anisotropic ellipsoidal YIG sphere within a microwave cavity, which enables the effective coupling between the microwave and magnon mode that is essential for the system's ability to sense weak magnetic fields. The YIG sphere, renowned for its superior magnetic properties, interacts with the electromagnetic fields inside the cavity. An independent semi-classical pump field drives the system by exciting the microwave cavity field mode. A weak external magnetic field, oriented along the x-axis, acts as the target signal for detection. This field disturbs the system, and its effect is observed through the magnon dynamics in the YIG sphere. Under the macroscopic spin density limit, where numerous spins act collectively, The Hamiltonian of the system, using Holstein-Primakoff transformations \cite{PhysRev.58.1098, PhysRevB.103.L100403, tiablikov2013methods}, reads in the rotating frame of the driving field as
\begin{eqnarray}\label{1}
	\hat{H^{\prime}}&=&\hbar\Delta _a \hat{a}^{\dagger} \hat{a}+\hbar \omega_{0} \hat{m}^{\dagger} \hat{m}-\frac{\hbar\eta_{m}}{2}\left(\hat{m}^{2}+\hat{m}^{\dagger 2}\right)+i\hbar E_L( \hat{a}^{\dagger}-\hat{a})\nonumber\\
	&+&\hbar g\left(\hat{m}\hat{a}^{\dagger}+\hat{m}^{\dagger}\hat{a}\right) -\hbar \lambda B_{ex}(t) (\hat{m}+\hat{m}^{\dagger}),
\end{eqnarray}
where $\Delta_{a} = \omega_{a} - \omega_{L}$ denotes the detuning between the cavity field and the driving microwave field, and $\hbar$ is the reduced Planck constant.  The first two terms of the Hamiltonian represent the free Hamiltonian of the microwave cavity field and the magnon mode in the YIG sphere, with $\omega_{a}$ and $\omega_{0}$ indicating the intrinsic frequencies of the cavity field and the magnon mode, respectively. The operators $\hat{a}$ and $\hat{m}$ $(\hat{a}^\dagger, \hat{m}^\dagger)$ correspond to the annihilation (creation) operators for microwave photons and magnons. The frequency $\omega_0$ of the magnon mode is adjustable by a bias magnetic field $B_b$ applied along the z-axis, with the gyromagnetic ratio $\gamma = 2\pi\times {28} ~\mathrm{GHz/ T}$ \cite{doi:10.1126/sciadv.1501286}. The third term describes parametric amplification due to the anisotropy of the ellipsoidal YIG sphere, with $\eta_m = \omega_0 -|\gamma| \mu_0 B_b $ representing the anisotropic coupling coefficient \cite{PhysRevLett.116.146601, PhysRevB.101.014416, PhysRevApplied.16.064008, PhysRevB.103.L100403},  with the details in Appendix ~\ref{appendix:A}. The fourth term corresponds to the semi-classical pump field driving the cavity field mode, where $E_L=\sqrt{2P_{L}\kappa_a/\hbar\omega_L}$ is the amplitude of the pump field with frequency $\omega_L$,  $P_L$ is the power of the microwave pump field and $\kappa_{a}$ is the dissipation rate of the microwave cavity field. The coupling strength between the cavity field and magnons is \( g = \gamma B_0\sqrt{5N}/2 \), with \( B_0 \) being the intensity of the microwave field. The fifth term accounts for the dipole-dipole interaction between the microwave photons and magnons. The last term represents the interaction between the weak external magnetic field \( B_{\text{ex}}(t) \) along the x-axis and the YIG sphere, with \( \lambda = \gamma \sqrt{5N}/2 \) being the coupling coefficient of the external field and the magnon mode. Here, \( N \) denotes the total number of spins in the system, which can be achieved experimentally, with \( N = 3.5 \times 10^{19} \) as reported in \cite{PhysRevLett.113.156401}. This Hamiltonian effectively describes the weak magnetic field sensing in an anisotropic cavity magnonic system.

Based on the Hamiltonian Eq. (\ref{1}), we can derive the quantum Heisenberg-Langevin equations for the system \cite{PhysRevLett.46.1, PhysRevA.31.3761} as
\begin{eqnarray}\label{2}
	\dot{\hat{m}}&=&-i\omega_0  \hat{m}-\frac{\kappa_m }{2} \hat m-ig \hat{a}+i\eta_{m}\hat{m}^{\dagger}+\sqrt{\kappa_m} \hat{m}_{\rm in}(t)+i\lambda B_{ex}(t), \nonumber\\
	\dot{\hat{a}}&=&-i\Delta_a \hat{a}-\frac{\kappa_a}{2}\hat{a}-ig \hat{m}+E_L+\sqrt{\kappa_a} \hat{a}_{\rm in}(t),
\end{eqnarray}
where $\kappa_m$ denotes the dissipation rate of the magnon mode, and $\hat{a}_{\rm in}(t)$, $\hat{m}_{\rm in}(t)$ represent the input noise operator of the cavity field and the magnon mode, respectively.  These equations capture the interplay between the coherent evolution governed by the Hamiltonian and the stochastic perturbations arising from environmental noise.  In the Markovian approximation, the noise properties of the input operators \(\hat{a}_{\rm in}(t)\) and \(\hat{m}_{\rm in}(t)\) are characterized by the following correlation functions
\begin{eqnarray}\label{3}
	&&\left\langle \hat a_{\rm in}(t) \hat a_{\rm in}^{\dagger}\left(t^{\prime}\right)\right\rangle = (\bar n_{a }+1)\delta\left(t-t^{\prime}\right), \notag\\
	&&\left\langle \hat a_{\rm in}^{\dagger}(t) \hat a_{\rm in}\left(t^{\prime}\right)\right\rangle = \bar n_{a}\delta\left(t-t^{\prime}\right), \notag\\
	&&\left\langle \hat m_{\rm in}(t) \hat m_{\rm in}^{\dagger}\left(t^{\prime}\right)\right\rangle = (\bar n_{m }+1)\delta\left(t-t^{\prime}\right), \notag\\
	&&\left\langle \hat m_{\rm in}^{\dagger}(t) \hat m_{\rm in}\left(t^{\prime}\right)\right\rangle = \bar n_{m}\delta\left(t-t^{\prime}\right).
\end{eqnarray}
Here,  $\bar n_{L(m)} = \left[\exp\left(\frac{\hbar \omega_{L (m)}}{k_B T}\right) - 1\right]^{-1}$ represents the average number of photons (magnons),  where $\omega_L$ and $\omega_0$ denote the frequencies of the driving microwave cavity field and magnon mode, respectively, $k_B$ is the Boltzmann constant, and $T$ is the temperature of the environment.  

Considering the strong coherent driving, all the operators can be written as the steady-state mean value plus its first order quantum fluctuation as $\hat a (\hat{m})=\bar{a}(\bar{m})+\delta {\hat{a}}(\delta {\hat{m}})$. Thus, one can obtain the Heisenberg-
Langevin's equation for the quantum fluctuation operators  as
\begin{eqnarray}\label{4}
	\delta\dot{\hat{a}}&=&-i\Delta_a \delta\hat{a}-\frac{\kappa_a}{2}\delta\hat{a}-ig \delta\hat{m}+\sqrt{\kappa_a} \hat{a}_{\rm in}(t),\nonumber\\
	\delta\dot{\hat{m}}&=&-i\omega_0  \delta\hat{m}-\frac{\kappa_m}{2} \delta\hat m+i\eta_{m}\delta\hat{m}^{\dagger}-ig \delta\hat{a}\notag\\&+&\sqrt{\kappa_m} \hat{m}_{\rm in}(t)+i\lambda B_{ex}(t).
\end{eqnarray}
Since the external magnetic field signal will be detected through the output of the microwave cavity field, we'd like to introduce the quadrature components  as $\delta\hat{X}_{a}=(\delta\hat{a}^{\dagger }+\delta\hat{a})/\sqrt{2}$, $\delta\hat{P}_{a}=(
\delta\hat{a}-\delta\hat{a}^{\dagger })/\sqrt{2}i$, $\delta\hat{X}_{m}=(
\delta\hat{m}+\delta\hat{m}^{\dagger })/\sqrt{2}i$, $\delta\hat{P}_{m}=(
\delta\hat{m}-\delta\hat{m}^{\dagger })/\sqrt{2}i$.  In this way,  the above Eq. (\ref{4}) can be rewritten as
\begin{widetext}
\begin{eqnarray}\label{5}
	\delta \dot{\hat{X}}_{m}&=&(\omega_0+\eta_{m}) \delta \hat{P}_{m}-\frac{\kappa_m}{2} \delta \hat{X}_{m}+g\delta {\hat{P}}_{a}+\sqrt{\kappa_m}\hat{X}_{m}^{\rm {in }},\notag \\ \delta \dot{\hat{P}}_{m}&=&(\eta_{m}-\omega_0) \delta \hat{X}_{m}-\frac{\kappa_m}{2} \delta \hat{P}_{m}-g\delta \hat{X}_{a}+\sqrt{\kappa_m}[\hat{P}_{m}^{\rm { in }}+ \tilde{B}_{ex}(t)], \notag\\ \delta \dot{\hat{X}}_{a}&=&\Delta_a \delta \hat{P}_{a}-\frac{\kappa_{a}}{2} \delta \hat{X}_{a}+g\delta \hat{P}_{m}+\sqrt{\kappa_{a}} \hat{X}_{a}^{\rm {in }},\notag \\ \delta \dot{\hat{P}}_{a}&=&-\Delta_a \delta \hat{X}_{a}- g \delta \hat{X}_{m}-\frac{\kappa_{a}}{2} \delta \hat{P}_{a}+\sqrt{\kappa_{a}} \hat{P}_{a}^{\rm {in }},
\end{eqnarray}
where $\hat{X}_{a}^{in}=(\hat{a}%
_{in}^{\dagger }+\hat{a}_{in})/\sqrt{2}$, $\hat{P}_{a}^{in}=(\hat{a}_{in}-%
\hat{a}_{in}^{\dagger })/\sqrt{2}i$, $\hat{X}_{m}^{\rm in}=(\hat{m}%
_{in}^{\dagger }+\hat{m}_{in})/\sqrt{2}$, $\hat{P}_{m}^{\rm in}=(\hat{m}_{in}-%
\hat{m}_{in}^{\dagger })/\sqrt{2}i$ are the input noise quadratures of the cavity field and magnon mode, $\tilde{B}_{ex}(t)=\sqrt{2/\kappa_{m}}\lambda B_{ex}(t)$ represents the rescaled magnetic field to be measured.
Since our system includes nonlinear terms, it is essential to analyze its stability in order to derive the threshold conditions necessary for linearization. From Eq. (\ref{5}), we can express the drift matrix as
\begin{equation}\label{eq:drift_matrix}
	A = 
	\begin{pmatrix}
		-\dfrac{\kappa_m}{2} & \omega_0 + \eta_m & 0 & g \\[6pt]
		\eta_m - \omega_0   & -\dfrac{\kappa_m}{2} & -g & 0 \\[6pt]
		0                   & g                   & -\dfrac{\kappa_a}{2} & \Delta_a \\[6pt]
		-\,g                & 0                   & -\Delta_a             & -\dfrac{\kappa_a}{2}
	\end{pmatrix}.
\end{equation}
According to the Routh–Hurwitz stability criterion \cite{PhysRevA.35.5288}, the characteristic polynomial of the drift matrix yields the following threshold conditions
\begin{align}
	R_3 > 0, R_3 R_2 - R_1 > 0, R_3 R_2 R_1 - R_1^2 - R_3^2 R_0 > 0. 
\end{align}
Here the coefficients \(R_i\) are given by
\begin{align}
	R_3 &= \kappa_a + \kappa_m, \nonumber\\[4pt]
	R_2 &= 2\,g^2 + \Delta_a^2
	+ \tfrac{1}{4}\bigl(-4\,\eta_m^2 + \kappa_a^2 + 4\,\kappa_a\kappa_m + \kappa_m^2\bigr)
	+ \omega_0^2, \nonumber\\[6pt]
	R_1 &= -\,\eta_m^2\,\kappa_a + \Delta_a^2\,\kappa_m + g^2(\kappa_a + \kappa_m)
	+ \tfrac{1}{4}\,\kappa_a\,\kappa_m(\kappa_a + \kappa_m)
	+ \kappa_a\,\omega_0^2, \nonumber\\[6pt]
	R_0 &= g^4
	+ \tfrac{1}{2}\,g^2\,\kappa_a\,\kappa_m
	- \tfrac{1}{16}\bigl(4\,\Delta_a^2 + \kappa_a^2\bigr)\bigl(4\,\eta_m^2 - \kappa_m^2\bigr).
	\label{eq:Ri}
\end{align}
In the following numerical analysis and discussion, we ensure that the stability conditions are satisfied.
To further address the impact of noise on the sensing system, we transform Eqs. (\ref{5}) into the frequency domain space using the Fourier transform ${\hat{O}(\omega)=\int_{-\infty}^{\infty}dtO(t)e^{i\omega t}}$. Utilizing the input-output relation for the phase quadrature component $\delta\hat{P}^{\rm out}_a=\sqrt{\kappa_a}\delta \hat{P}_a-\hat{P}^{\rm in}_a$,  one can obtain  the phase quadrature of the cavity output field as
	\begin{eqnarray}\label{7}
		\delta\hat{P}^{\rm{out}}_a(\omega)&=&K_1(\omega)\hat{X}^{ \rm{in}}_a(\omega)+K_2(\omega)\hat{P}^{ \rm{in}}_a(\omega)+K_3(\omega)\hat{X}^{\rm{in}}_m(\omega)+K_4(\omega)(\hat{P}^{\rm{in}}_m(\omega)+\tilde{B}_{ex}(\omega)),
	\end{eqnarray}
	where
	\begin{align}\label{8}
		{K_1}(\omega)&= \frac{4 \kappa_a \left( 4 g^2 (\omega_0 + \eta_m) + \Delta_a \left(-4 (\omega_0 - \eta_m)(\omega_0 + \eta_m) + (2\omega + i \kappa_m)^2 \right) \right)}{
			16 g^4 + \left(-4 \Delta_a^2 + (2\omega + i \kappa_a)^2 \right) \left(-4 (\omega_0 - \eta_m)(\omega_0 + \eta_m) + (2\omega + i \kappa_m)^2 \right) 
			- 8 g^2 \left( 4\omega^2 + 4\omega_0 \Delta_a - \kappa_a \kappa_m + 2i \omega (\kappa_a + \kappa_m) \right)},
		\nonumber\\
		{K_2}(\omega)&=
		\frac{
			-16 g^4 - (4\omega^2 - 4\Delta_a^2 + \kappa_a^2)(-4(\omega_0 - \eta_m)(\omega_0 + \eta_m) + (2\omega + i\kappa_m)^2) 
			+ 16 g^2 (2\omega^2 + 2\omega_0\Delta_a + i\omega\kappa_m)
		}{
			16 g^4 + \left(-4\Delta_a^2 + (2\omega + i\kappa_a)^2\right)\left(-4(\omega_0 - \eta_m)(\omega_0 + \eta_m) + (2\omega + i\kappa_m)^2\right) 
			- 8 g^2 \left(4\omega^2 + 4\omega_0\Delta_a - \kappa_a\kappa_m + 2i\omega(\kappa_a + \kappa_m)\right)
		},\nonumber\\
		{K_3}(\omega)&=
		\frac{
			-4g \sqrt{\kappa_a} \sqrt{\kappa_m} \left( 4g^2 - 4\omega^2 - 4\omega_0\Delta_a + 4\Delta_a\eta_m + \kappa_a\kappa_m - 2i \omega (\kappa_a + \kappa_m) \right)
		}{
			16g^4 + \left(-4\Delta_a^2 + (2\omega + i\kappa_a)^2 \right) \left(-4(\omega_0 - \eta_m)(\omega_0 + \eta_m) + (2\omega + i\kappa_m)^2 \right)
			- 8g^2 \left( 4\omega^2 + 4\omega_0\Delta_a - \kappa_a\kappa_m + 2i \omega (\kappa_a + \kappa_m) \right)
		},\nonumber\\
		{K_4}(\omega)&=	\frac{-
			8g \sqrt{\kappa_a} \sqrt{\kappa_m} \left( -2i \omega (\omega_0 + \Delta_a + \eta_m) + (\omega_0 + \eta_m) \kappa_a + \Delta_a \kappa_m \right)
		}{
			16g^4 + \left( -4\Delta_a^2 + (2\omega + i\kappa_a)^2 \right) \left( -4(\omega_0 - \eta_m)(\omega_0 + \eta_m) + (2\omega + i\kappa_m)^2 \right)
			- 8g^2 \left( 4\omega^2 + 4\omega_0\Delta_a - \kappa_a\kappa_m + 2i\omega (\kappa_a + \kappa_m) \right)
		}.
	\end{align}
We can find that the output phase quadrature \(\delta\hat{P}^{\rm{out}}_a(\omega)\) is a linear combination of the cavity input noise quadratures (\(\hat{X}_a^{\rm{in}}(\omega)\), \(\hat{P}_a^{\rm{in}}(\omega)\)), the magnon input noise quadratures (\(\hat{X}_m^{\rm{in}}(\omega)\), \(\hat{P}_m^{\rm{in}}(\omega)\)), and the external magnetic field signal \(\tilde{B}_{\rm ex}(\omega)\). The coefficients \(K_1(\omega)\), \(K_2(\omega)\), \(K_3(\omega)\), and \(K_4(\omega)\) represent the frequency-dependent contributions of these components, determined by the cavity-magnon coupling strength \(g\), the cavity-drive field detuning \(\Delta_a\), the magnon frequency \(\omega_0\), their dissipation rates (\(\kappa_a\), \(\kappa_m\)), and the anisotropic parameter ($\eta_m$). Specifically, \(K_4(\omega)\) directly couples the magnetic field to the output, enabling its detection via the cavity field output.
\end{widetext}	
\section{\label{sec3} Weak magnetic field sensing using homodyne output spectrum analysis}

In homodyne detection, by adjusting the phase \(\varphi\) of the local oscillator, we can measure any generalized quadrature component. The detected signal photocurrent operature \(\hat{I}_{\text{De}}\) is proportional to the generalized quadrature component \(\hat{X}_{a,\text{out}}^{\varphi}\), and
the generalized quadrature component \(\hat{X}_{a,\text{out}}^{\varphi}\) is given by
\begin{eqnarray}
\hat{X}_{a,\text{out}}^{\varphi} = \cos(\varphi)\delta \hat{X}^{\rm{out}}_a + \sin(\varphi) \delta \hat{P}^{\rm{out}}_a
\end{eqnarray}
in our subsequent discussion, we focus on the phase quadrature of the cavity field with \(\varphi = \pi/2\), simplifying the expression to \(X_{a,\text{out}}^{\pi/2} = \delta \hat{P}^{\rm{out}}_a\). Homodyne detection is crucial in weak magnetic field sensing systems, enabling effective phase output spectrum calculations, expressed as \cite{PhysRevA.103.062605, PhysRevA.109.023709, PhysRevA.100.023815,Leonhardt1997}
	\begin{align}\label{9}
		S_{{P\mathrm{out}}}(\omega)=\frac{1}{4\pi}\int d\omega^{\prime}e^{i(\omega+\omega^{\prime})t}\left[C_P(\omega,\omega')+C_P(\omega',\omega)\right]
	\end{align}
	with $C_P(\omega,\omega')=\langle\delta \hat{P}^{\rm{out}}_a(\omega)\delta \hat{P}^{\rm{out}}_a(\omega^{\prime})\rangle$.
	For our system, we obtain
	\begin{align}\label{10}
		S_{P{\rm{out}}}(\omega)&=(\bar{n}_a+\frac{1}{2})[\vert {K_1}(\omega)\vert ^2+\vert {K_2}(\omega)\vert ^2]\notag\\&+(\bar{n}_m+\frac{1}{2})[\vert {K_3}(\omega)\vert ^2+\vert {A_4}(\omega)\vert ^2]+\vert {K_4}(\omega)\vert ^2S_{\tilde{B}_{ex}}(\omega).
	\end{align}
Here, $S_{\tilde{B}_{ex}}$ represents the signal spectral density corresponding to the external magnetic field.
To better understand the noise suppression performance, one can express the output power spectrum as
	\begin{align}\label{11}
		S_{P{\rm{out}}}(\omega)=A_m(\omega)[N_{\rm{qn}}(\omega)+N_{mth}(\omega)+S_{\tilde{B}_{ex}}(\omega)],
	\end{align}
	where $ A_m(\omega) $, $ N_{\rm{qn}}(\omega) $, and $ N_{mth}(\omega) $ represent the system's response to the detected magnetic field, additional quantum noise of the cavity field, and magnon input thermal noise, respectively. These are explicitly given as
	\begin{align}\label{12}
		&A_m(\omega)=\vert{K_4}(\omega)\vert ^2, \notag\\
		&N_{\rm{\rm{qn}}}(\omega)=(\bar{n}_a+\frac{1}{2})\frac{ \vert {K_1}(\omega)\vert ^2+ \vert {K_2}(\omega)\vert ^2}{\vert {K_4}(\omega)\vert ^2},\notag\\
		&N_{mth}(\omega)=(\bar{n}_m+\frac{1}{2})[\frac{\vert {K_3}(\omega)\vert ^2}{\vert {K_4}(\omega)\vert ^2}+1].
	\end{align}
	\begin{figure*}
	\includegraphics[width=18cm,height=8.0cm]{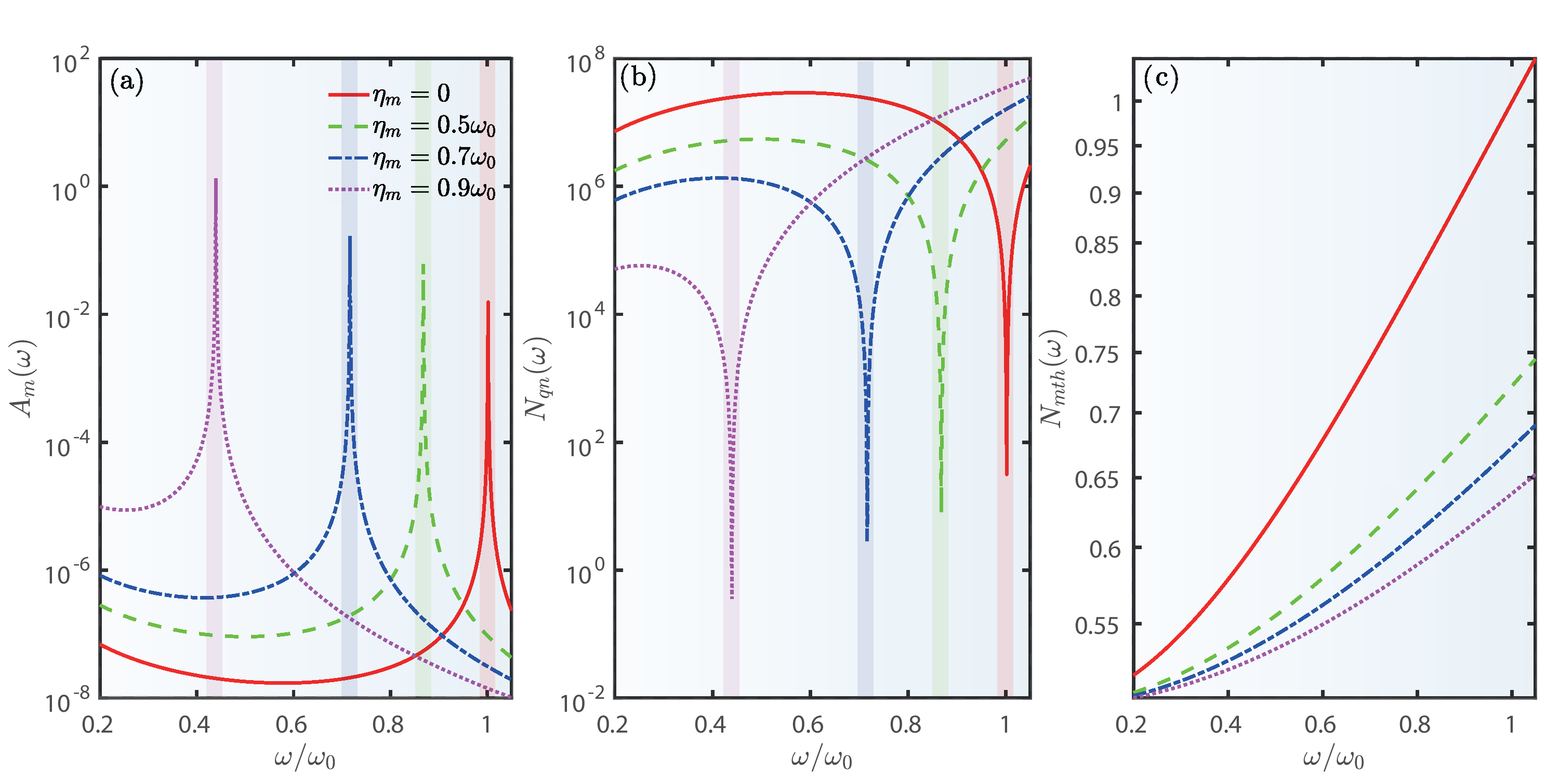}
	\caption{(a) The response \(A_{\rm m}(\omega)\) of the weak magnetic field sensing system as a function of the normalized frequency \(\omega/\omega_0\), shown for various anisotropy parameters: \(\eta_m = 0\) (solid red line), \(\eta_m = 0.5\omega_0\) (dashed green line), \(\eta_m = 0.7\omega_0\) (dash-dotted blue line), and \(\eta_m = 0.9\omega_0\) (dotted magenta line). (b) The additional quantum noise of the cavity field \(N_{\rm qn}(\omega)\) as a function of the normalized frequency \(\omega/\omega_0\), plotted under different anisotropy parameters: \(\eta_m = 0\) (solid red line), \(\eta_m = 0.5\omega_0\) (dashed green line), \(\eta_m = 0.7\omega_0\) (dash-dotted blue line), and \(\eta_m = 0.9\omega_0\) (dotted magenta line). (c) The magnon mode thermal input noise \(N_{\rm mth}(\omega)\) as a function of the normalized frequency \(\omega/\omega_0\), also shown for different anisotropy parameters: \(\eta_m = 0\) (solid red line), \(\eta_m = 0.5\omega_0\) (dashed green line), \(\eta_m = 0.7\omega_0\) (dash-dotted blue line), and \(\eta_m = 0.9\omega_0\) (dotted magenta line). The initial environmental temperature is set at 5 mK, and the detuning between the cavity field and the driving microwave field is \(\Delta_a = 0\).}
		\label{Fig2}
\end{figure*}

\begin{figure*}
	\includegraphics[width=18cm,height=8.0cm]{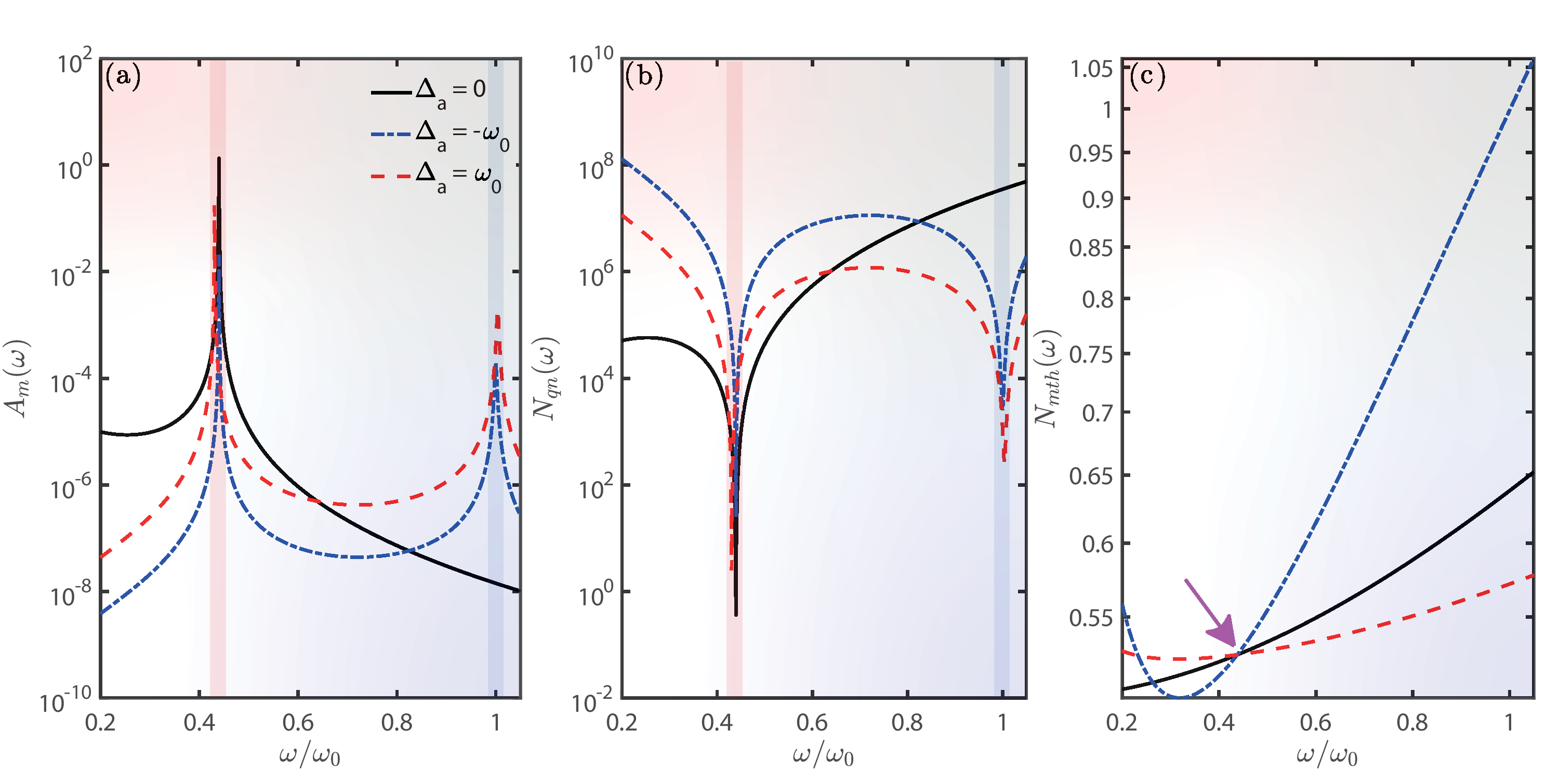}
	\caption{(a) The response \(A_{\rm m}(\omega)\) of the weak magnetic field sensing system as a function of the normalized frequency \(\omega/\omega_0\), shown for various detuning parameters: \(\Delta_a = 0\) (solid black line), \(\Delta_a = -\omega_0\) (dash-dotted blue line), and \(\Delta_a = \omega_0\) (dashed red line). (b) The additional quantum noise of the cavity field \(N_{\rm qn}(\omega)\) as a function of the normalized frequency \(\omega/\omega_0\), plotted under the same detuning conditions: \(\Delta_a = 0\) (solid black line), \(\Delta_a = -\omega_0\) (dash-dotted blue line), and \(\Delta_a = \omega_0\) (dashed red line). (c) The magnon mode thermal input noise \(N_{\rm mth}(\omega)\) as a function of the normalized frequency \(\omega/\omega_0\), also shown for the same detuning conditions: \(\Delta_a = 0\) (solid black line), \(\Delta_a = -\omega_0\) (dash-dotted blue line), and \(\Delta_a = \omega_0\) (dashed red line). The initial environmental temperature is set at 5 mK, and the anisotropy parameter is fixed at \(\eta_m = 0.9\omega_0\).}
	\label{Fig3}
\end{figure*}
	\begin{figure*}
\centering\includegraphics[width=18cm,height=8.0cm]{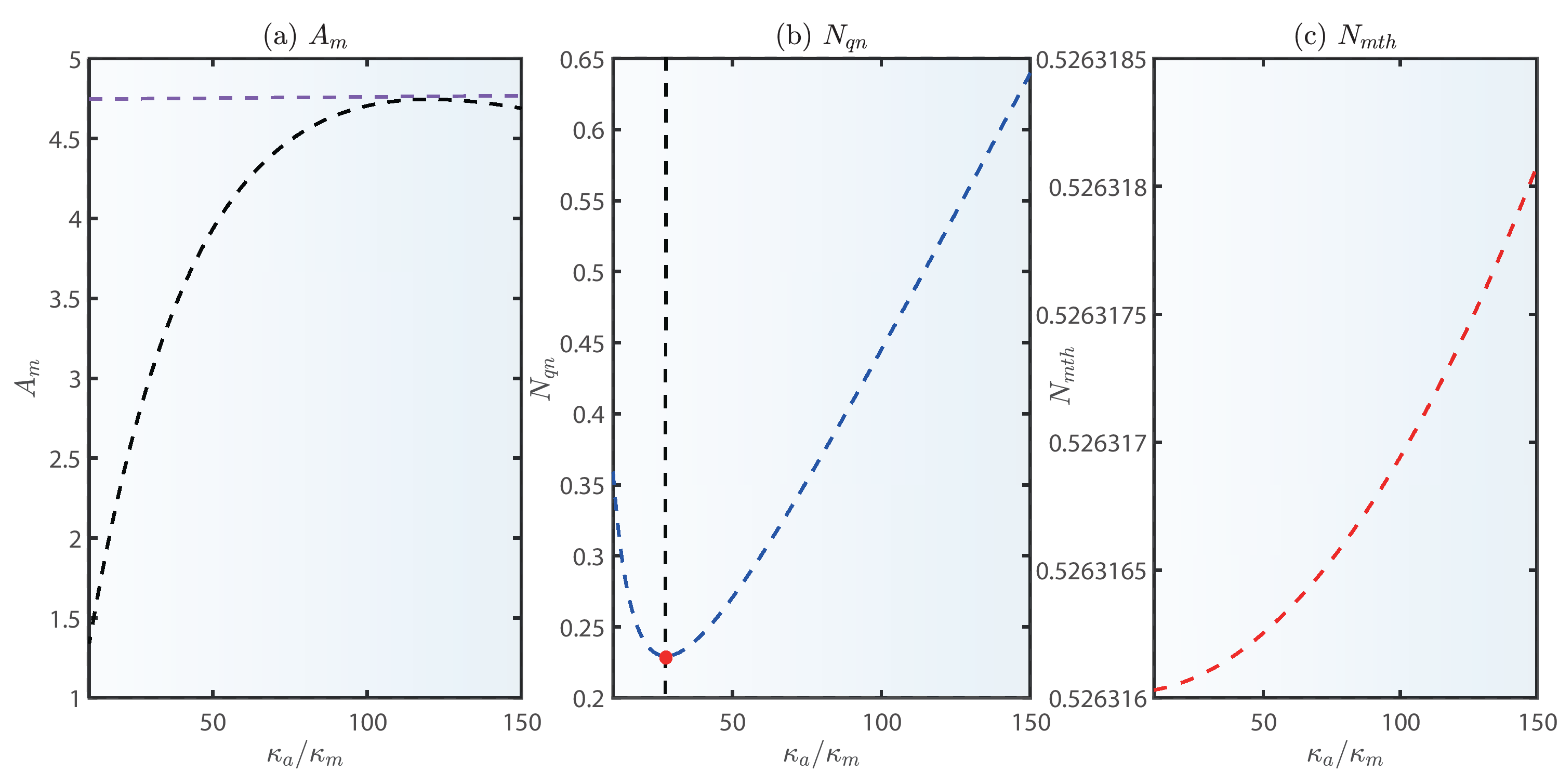}
		\caption{(a) The response $A_{\rm m}(\omega)$ of the weak magnetic field sensing system at the optimal detection frequency as a function of the normalized cavity field dissipation $\kappa_a / \kappa_m$. (b) The additional quantum noise of the cavity field \(N_{qn}(\omega)\) at the optimal detection frequency as a function of the normalized cavity field dissipation \(\kappa_a / \kappa_m\). The noise increases monotonically with dissipation but exhibits a minimum within a small range (i.e., the optimal noise point). (c) Thermal input noise at the optimal detection frequency as a function of the normalized cavity field dissipation \(\kappa_a / \kappa_m\). The initial environmental temperature is set at 5 mK, and the anisotropy parameter is fixed at \(\eta_m = 0.9\omega_0\).}
		\label{Fig4}
	\end{figure*}

\begin{figure*}
	\centering\includegraphics[width=18cm,height=8.0cm]{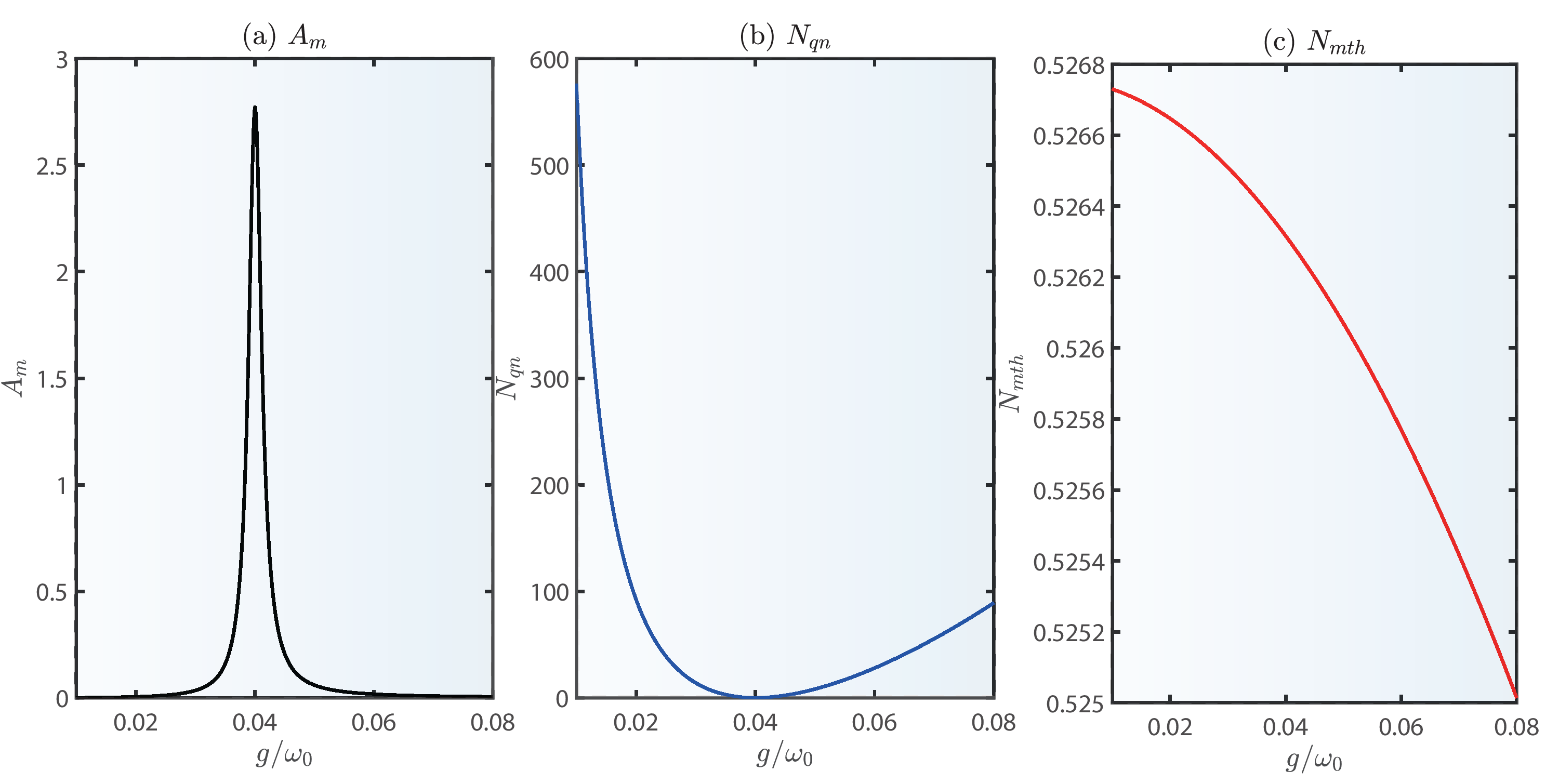}
	\caption{(a) The response \(A_m(\omega)\) of the weak magnetic field sensing system at the optimal detection frequency as a function of the normalized coupling strength \(g / \omega_0\). The response exhibits a sharp peak at a specific coupling strength. (b) The additional quantum noise of the cavity field \(N_{qn}(\omega)\) at the optimal detection frequency as a function of the normalized coupling strength \(g / \omega_0\). The noise decreases with increasing coupling strength, reaching a minimum before rising again. (c) The thermal input noise \(N_{mth}(\omega)\) at the optimal detection frequency as a function of the normalized coupling strength \(g / \omega_0\). The thermal noise decreases monotonically as the coupling strength increases. The initial environmental temperature is set to 5 mK, with the cavity field dissipation \(\kappa_a = 26\kappa_m\).}
	\label{Fig5}
\end{figure*}
	To better understand the impact of various mechanisms on noise suppression, we conducted numerical simulations of several key sensing performance indicators at low temperatures, specifically at $5\, \rm mK$. For this study, we utilized feasible parameters  \cite{PhysRevLett.113.156401,PhysRevLett.111.127003, PhysRevB.103.L100403, PhysRevA.103.062605}, such as  $\omega_{0}/2\pi = \omega_{a}/2\pi = 25\, \rm {GHz}$, $\lambda/2\pi =  1.85 \times 10^{20} \rm {Hz}/T$, $g =100\kappa_m$, $\kappa_{m}/2\pi = 10\, \rm {MHz}$, and $\kappa_{a}/2\pi = 100\, \rm {MHz}$.
\subsection{Sensing Performance of the Probe Main Axis Along the y-direction}
	In this subsection, we first investigate the case where the semi-major axis of the ellipsoidal magnet aligns with the y-axis (i.e., the anisotropy coefficient (\(\eta_m > 0\))), as shown in Fig. \ref{Fig1}, to analyze the sensing performance of the system.
	
Fig. \ref{Fig2} illustrates the dependence of three key performance indicators of the weak magnetic field sensing system—the response $A_m$, the cavity field quantum noise $N_{qn}$, and the probe input thermal noise $N_{mth}$ —on the normalized frequency \(\omega / \omega_0\). These indicators comprehensively analyze the impact of varying the anisotropic parameter \(\eta_m\) on system performance. Fig. \ref{Fig2}(a) shows the variation of the response function \(A_m(\omega)\) with frequency. When \(\eta_m = 0\), the system's response peaks are near \(\omega / \omega_0 = 1\). As \(\eta_m\) increases, the response peak grows significantly, indicating enhanced response to external signals while shifting to lower frequencies. For example, with \(\eta_m = 0.9 \omega_0\), the peak occurs at \(\omega / \omega_0 \approx 0.44\) and exceeds unity, indicating signal amplification. This demonstrates that squeezing-induced anisotropy enhances the response magnitude while lowering the optimal detection frequency.
Fig. \ref{Fig2}(b) shows the cavity field quantum noise \(N_{qn}(\omega)\). When \(\eta_m = 0\), quantum noise suppression is limited, resulting in a distinct valley at \(\omega / \omega_0 = 1\). As \(\eta_m\) increases, the depth of the noise valley increases, and the minimum noise shifts to align with the optimal detection frequency observed in Fig. \ref{Fig2}(a). This deepening of the valley also reflects the system's ability to suppress cavity field quantum noise. For example, with \(\eta_m = 0.9 \omega_0\), the minimum noise occurs at approximately \(\omega / \omega_0 \approx 0.44\), which is about two orders of magnitude lower compared to \(\eta_m = 0\).
Fig. \ref{Fig2}(c) presents the variation of magnon probe input thermal noise \(N_{mth}(\omega)\). When \(\eta_m = 0\), the thermal input noise of the magnon probe increases almost linearly with frequency. The overall trend remains consistent across different \(\eta_m\) values, but absolute noise levels decrease as \(\eta_m\) increases. For example, with \(\eta_m = 0.9 \omega_0\), thermal noise is significantly reduced across the entire frequency range, reflecting a global thermal input noise reduction due to space anisotropy, independent of specific frequencies. This reduction complements the frequency-dependent improvements observed in Fig. \ref{Fig2}(a) and (b), further boosting overall system performance. Therefore, increasing \(\eta_m\) enhances the system's response, shifts the optimal detection frequency to lower values, suppresses quantum noise at the optimal frequency, and reduces thermal input noise across all frequencies. These findings underscore the critical role of  geometric anisotropy in optimizing weak magnetic field sensing performance.
	
To analyze the impact of cavity-drive field detuning on the sensing performance of the system, the anisotropy parameter was fixed at \(\eta_m = 0.9 \omega_0\). Fig. \ref{Fig3} presents the system's responses under three distinct detuning conditions, highlighting the variations in cavity field quantum noise and probe input thermal noise. In Fig. \ref{Fig3}(a), while two peaks emerge under detuning conditions (\(\Delta_a = \pm \omega_0\)), neither surpasses the peak value observed at the optimal detection frequency under the zero-detuning condition (\(\Delta_a = 0\)), as indicated by the solid black curve. This result suggests that the signal amplification capability is maximized under zero detuning. For example, at \(\omega / \omega_0 \approx 0.44\), the black curve (zero detuning) exhibits a higher amplitude compared to the blue dash-dotted curve (\(\Delta_a = -\omega_0\)) and the red dashed curve (\(\Delta_a = \omega_0\)), underscoring the superior amplification performance under zero-detuning conditions. Fig.  \ref{Fig3}(b) illustrates the cavity field quantum noise \(N_{qn}(\omega)\) under the same detuning conditions. Under detuning conditions (\(\Delta_a = \pm \omega_0\)), two noise valleys are observed, suggesting that detuning influences detection near the magnon resonance frequency ($\omega\approxeq\omega_0$). Notably, positive detuning (\(\Delta_a = \omega_0\)) results in better noise suppression than negative detuning (\(\Delta_a = -\omega_0\)), as evidenced by the smaller noise valley in the red dashed curve compared to the blue dash-dotted curve. However, the noise suppression under detuning conditions remains relatively weak compared to zero detuning. At the effective detection frequency (\(\omega / \omega_0 \approx 0.44\)), the quantum noise of the cavity field under zero detuning (solid black line) is significantly reduced, demonstrating exceptional sensing performance. Fig. \ref{Fig3}(c) presents the variation of probe input thermal noise \(N_{mth}(\omega)\). Interestingly, at the effective detection frequency (\(\omega / \omega_0 \approx 0.44\)), the thermal noise curves for all three detuning conditions converge at a single point, indicating that detuning does not influence probe input thermal noise at the optimal detection frequency. However, near the magnon resonance frequency, the thermal noise under red detuning (\(\Delta_a = \omega_0\)) is significantly lower than that under zero detuning. This suggests thermal noise suppression can be enhanced by adjusting the detuning to the red-detuned regime. For instance, at higher normalized frequencies (\(\omega / \omega_0 > 0.5\)), the red dashed curve (\(\Delta_a = \omega_0\)) exhibits significantly lower thermal noise compared to the black solid curve (zero detuning). This demonstrates that magnon thermal noise (\(\omega / \omega_0 > 0.5\)) can be effectively mitigated by tuning the system to a red-detuned condition. According to the Hamiltonian in Eq. (\ref{1}), we implement a beam-splitter-type coupling between the cavity field and the magnon mode. This type of coupling has a stronger effect under positive detuning (low-frequency resonance) compared to negative detuning (high-frequency anti-resonance). As a result, positive detuning facilitates more efficient extraction of information from the external magnetic field, leading to better noise suppression. Additionally, the zero-detuning condition ensures effective signal amplification and cavity field quantum noise suppression, making it particularly suitable for high-precision sensing applications. In contrast, red detuning enhances magnon input thermal noise suppression, especially at higher frequencies. Zero detuning can be used to optimize signal amplification based on specific sensing application requirements, while red detuning can be leveraged to minimize thermal noise, enabling highly sensitive detection of weak magnetic fields.

Next, we fix the detection frequency of the system at the effective detection point and investigate the variation of three sensing metrics as a function of the dissipation ratio between the cavity field mode and the magnon mode, as shown in Fig. \ref{Fig4}. From Fig. \ref{Fig4}(a), it can be observed that the system's response amplitude \(A_m(\omega)\) increases rapidly with the dissipation ratio \(\kappa_a/\kappa_m\) in the low dissipation regime and gradually saturates at higher values. For example, at \(\kappa_a/\kappa_m = 15\), the response amplitude is approximately \(A_m \approx 2\), while at \(\kappa_a/\kappa_m = 50\), it increases to \(A_m \approx 4\). Beyond \(\kappa_a/\kappa_m \approx 100\), the response begins to level off, approaching the saturation limit indicated by the purple dashed line at \(A_m \approx 4.74\). This consistent signal amplification suggests that a certain level of cavity field dissipation enhances the system's ability to detect weak magnetic field signals. However, excessive dissipation can introduce more noise, which negatively impacts the improvement of sensing performance.
Fig. \ref{Fig4}(b) reveals a particularly intriguing behavior in the additional cavity field quantum noise \(N_{qm}(\omega)\). Unlike the response amplitude, \(N_{qm}\) does not monotonically increase with the dissipation ratio. Instead, it exhibits a non-monotonic trend. Starting from \(\kappa_a/\kappa_m = 10\), the quantum noise initially decreases, reaching its minimum at approximately \(\kappa_a/\kappa_m = 26\) (marked by the red dot). At this optimal point, the quantum noise is minimized to \(N_{qm} \approx 0.23\), balancing the trade-off between extracting information from the system and reading it through external dissipation channels. However, as \(\kappa_a/\kappa_m\) increases beyond 50, the quantum noise starts to rise again, reaching \(N_{qm} \approx 0.64\) at \(\kappa_a/\kappa_m = 150\). This non-monotonic behavior underscores the importance of tuning dissipation to achieve optimal system sensing performance.
Fig. \ref{Fig4}(c) depicts the magnon thermal input noise \(N_{mth}(\omega)\) as a function of the dissipation ratio. Unlike the other two metrics, the thermal input noise increases monotonically with \(\kappa_a/\kappa_m\). For instance, at \(\kappa_a/\kappa_m = 50\), the thermal noise is \(N_{mth} \approx 0.526316\), and as the dissipation ratio increases to \(\kappa_a/\kappa_m = 150\), the noise rises to approximately \(N_{mth} \approx 0.526318\). Despite the monotonic growth, the change in input thermal noise remains within the same order of magnitude, indicating that input thermal noise is relatively insensitive to dissipation changes. This observation is critical because it suggests that the dissipation ratio can be adjusted to minimize the additional quantum noise without significantly increasing the input thermal noise. Thus, the optimal dissipation condition near \(\kappa_a/\kappa_m \approx26\) provides a practical guideline for reducing quantum noise while maintaining manageable thermal input noise levels.
\begin{figure}
	\centering\includegraphics[width=9.5cm,height=7cm]{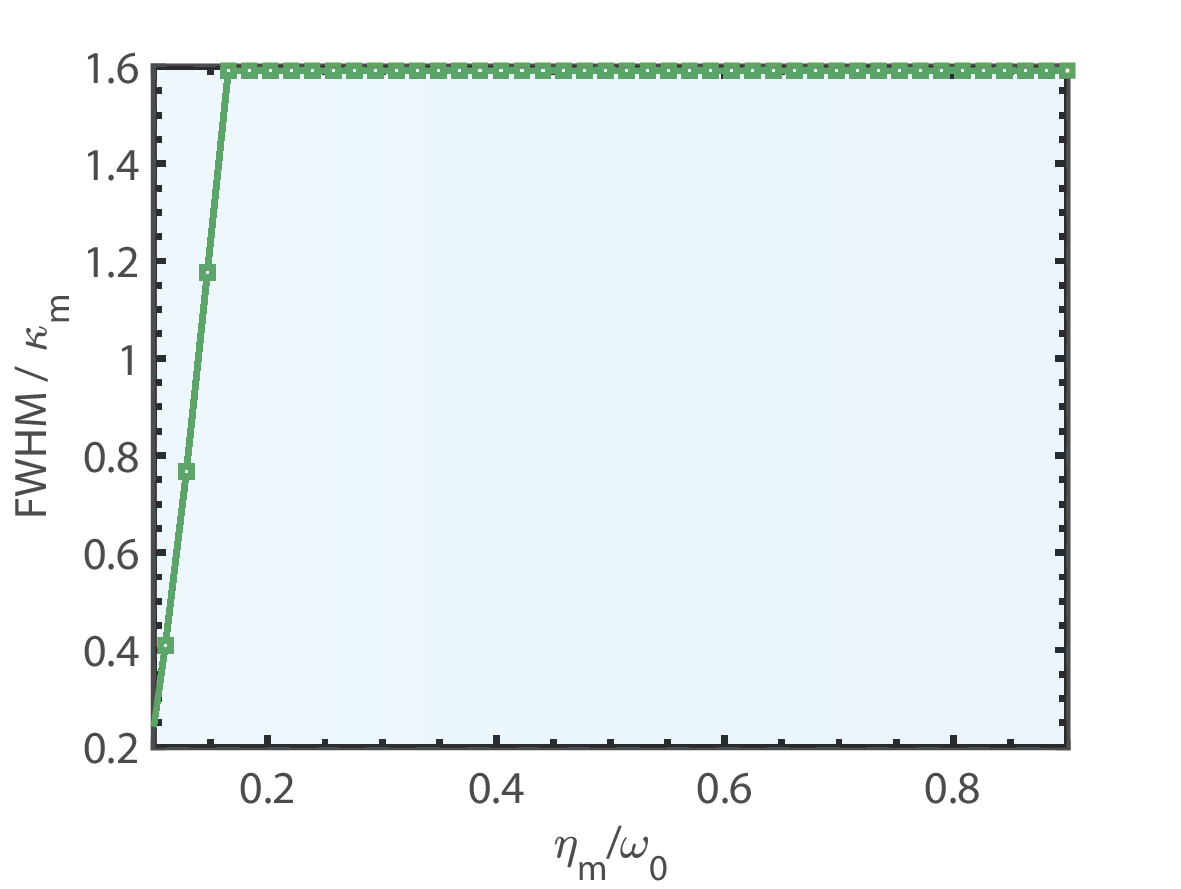}
	\caption{The response full width at half maximum (FWHM) as a function of the normalized anisotropy parameter. Other parameters are the same as Fig. \ref{Fig4}}.
	\label{Fig6}
\end{figure}
\begin{figure}
	\centering\includegraphics[width=9.4cm,height=11.5cm]{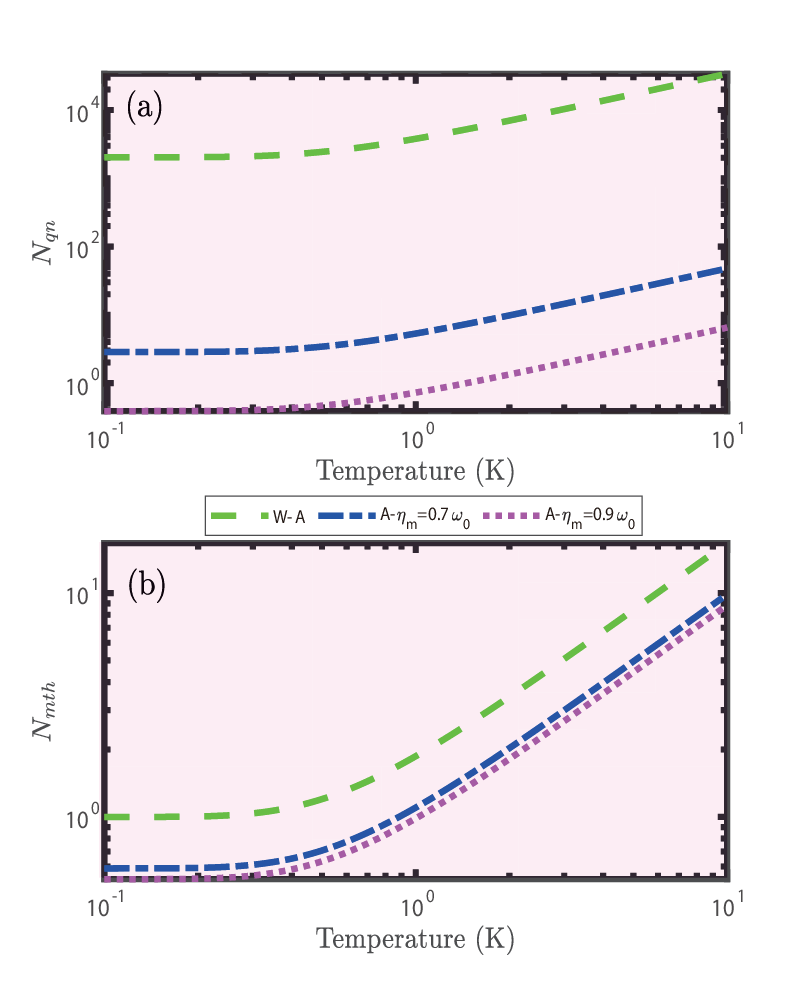}
	\caption{(a) The additional quantum noise of the cavity field \(N_{qn}(\omega)\) at the optimal detection frequency as a function of temperature. The different curves correspond to different values of the anisotropy parameter, namely \(A - \eta_m = 0.7 \omega_0\) (blue dashed line), \(A - \eta_m = 0.9 \omega_0\) (purple dotted line), and W-A (green dashed line), representing the spherical YIG sphere configuration. Here, "A" refers to the case where there is geometric anisotropy, while "W-A" refers to the case where there is no anisotropy. (b) The thermal magnon input noise \(N_{mth}(\omega)\) at the optimal detection frequency as a function of temperature.}
	\label{Fig7}
\end{figure}
	\begin{figure*}
	\centering\includegraphics[width=18cm,height=8.0cm]{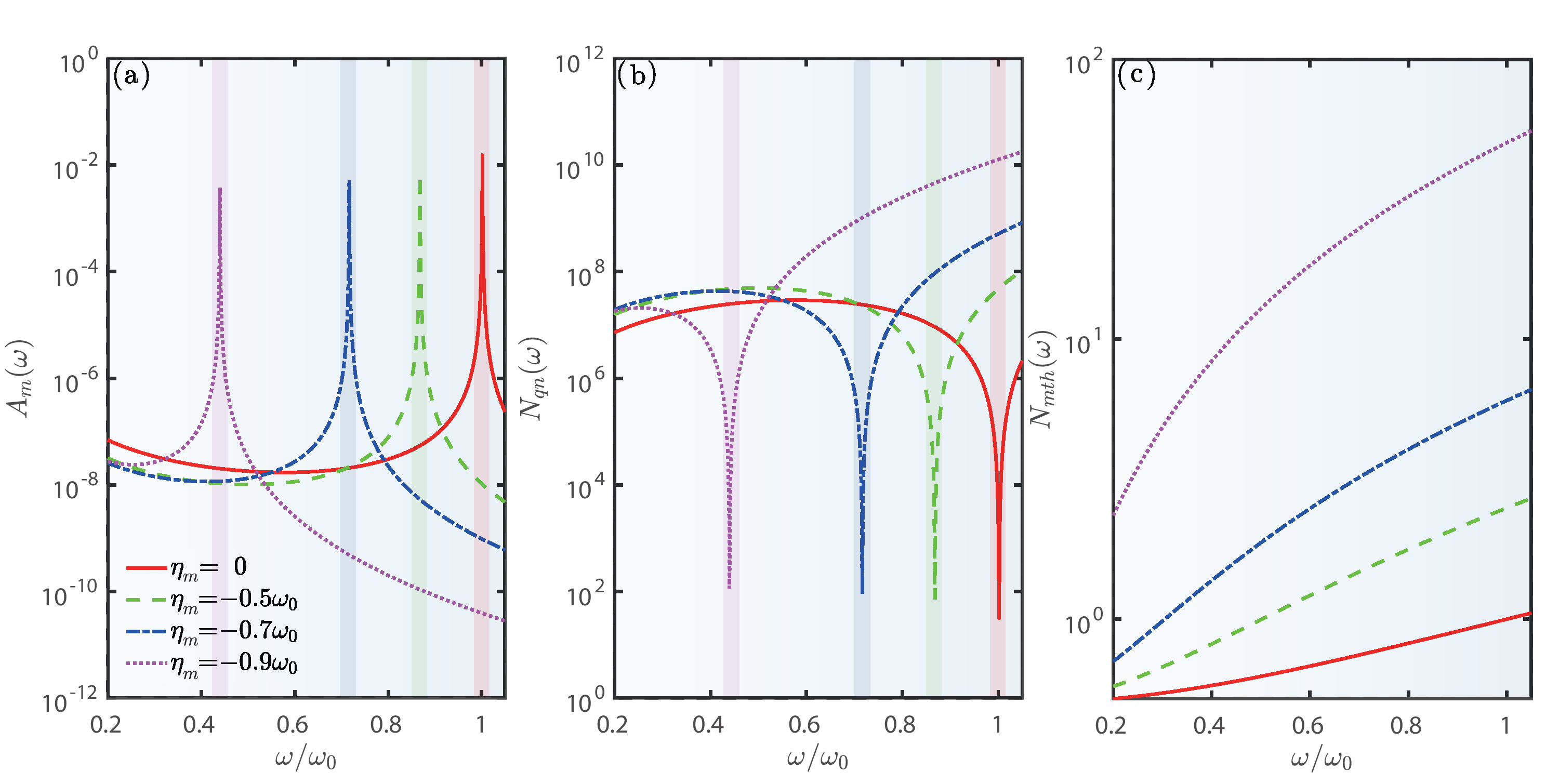}
	\caption{(a) The response \(A_{\rm m}(\omega)\) of the weak magnetic field sensing system as a function of the normalized frequency \(\omega/\omega_0\), shown for various anisotropy parameters: \(\eta_m = 0\) (solid red line), \(\eta_m = 0.5\omega_0\) (dashed green line), \(\eta_m = 0.7\omega_0\) (dash-dotted blue line), and \(\eta_m = 0.9\omega_0\) (dotted magenta line). (b) TThe additional quantum noise of the cavity field \(N_{\rm qn}(\omega)\) as a function of the normalized frequency \(\omega/\omega_0\), plotted under different anisotropy parameters: \(\eta_m = 0\) (solid red line), \(\eta_m = -0.5\omega_0\) (dashed green line), \(\eta_m = -0.7\omega_0\) (dash-dotted blue line), and \(\eta_m = -0.9\omega_0\) (dotted magenta line). (c) The thermal magnon input noise \(N_{\rm mth}(\omega)\) as a function of the normalized frequency \(\omega/\omega_0\), also shown for different anisotropy parameters: \(\eta_m = 0\) (solid red line), \(\eta_m = -0.5\omega_0\) (dashed green line), \(\eta_m = -0.7\omega_0\) (dash-dotted blue line), and \(\eta_m =-0.9\omega_0\) (dotted magenta line). The initial environmental temperature is set at 5 mK, and the detuning between the cavity field and the driving microwave field is \(\Delta_a = 0\).}
	\label{Fig8}
\end{figure*}

The cavity-magnon coupling strength is another significant parameter influencing the system's response and additional noise. Fig. \ref{Fig5} illustrates the variation of the system's response \(A_m(\omega)\), cavity field quantum noise \(N_{qn}(\omega)\), and probe input thermal noise \(N_{mth}(\omega)\) as a function of the normalized cavity-magnon coupling strength \(g/\omega_0\) under the optimal dissipation conditions identified in Fig. \ref{Fig4}(b). These results reveal the impact of coupling strength on system performance, including the system's response, additional quantum noise suppression, and probe input thermal noise reduction.
\textcolor{red}{As shown in Fig. \ref{Fig5}(a), the system's response reaches a peak at \(g/\omega_0 \approx 0.04\), indicating that the system operates in the coherent strong coupling regime \cite{PhysRevLett.111.127003,PhysRevLett.113.083603,doi:10.1126/sciadv.1501286}. Coherent strong coupling significantly enhances the energy exchange efficiency between cavity photons and magnons, enabling the system to detect weak magnetic field signals with high precision. The appearance of the response peak reflects a balance between energy exchange and dissipation, resulting in a marked improvement in the system's response.}
Fig. \ref{Fig5}(b) shows that quantum noise decreases significantly with increasing \(g/\omega_0\), reaching a minimum near \(g/\omega_0 \approx 0.04\), and then slightly increases. This demonstrates that coherent strong coupling enhances the system's response and minimizes quantum noise, effectively improving the signal extraction capability, which is critical for achieving high-precision sensing.
In contrast, Fig. \ref{Fig5}(c) shows that input magnon mode thermal noise decreases monotonically as \(g/\omega_0\) increases. However, the reduction is relatively smaller than the quantum noise of the cavity field. From the previous analysis, it is evident that the approach of geometric modulation anisotropy has the potential to amplify magnetic field signals. However, whether the response bandwidth increases with enhanced anisotropy remains an interesting question. To explore this, we quantified the response bandwidth using the full width at half maximum (FWHM) of the magnetic field response, as shown in Fig. \ref{Fig6}. The results indicate that as anisotropy increases, the bandwidth also increases accordingly, demonstrating the advantage of this approach. However, with further increase, the effect saturates, with the bandwidth essentially reaching a plateau around $1.6\kappa_m$. This may be attributed to the fact that when the anisotropy parameter becomes sufficiently large, the local response within the material or structure gradually saturates, making the system less sensitive to further changes in the anisotropy parameter. As a result, the bandwidth expansion tends to stabilize.

Aside from certain internal parameters that affect the system's sensing performance, external parameters, such as temperature, also play a role in influencing the system. Since the sensing analysis we previously conducted was performed at a temperature of 5 mK, we aimed to illustrate the sensing system's robustness concerning temperature variations. We compared configurations with and without anisotropy to achieve this, as shown in Fig. \ref{Fig7}. From the comparison, it is evident that when \(\eta_m \neq 0\), both the additional cavity field quantum noise (Fig. \ref{Fig7}(a)) and the thermal input noise (Fig. \ref{Fig7}(b)) display strong robustness against fluctuations in temperature. Specifically, when the system incorporates anisotropy, the noise suppression performance demonstrates reduced sensitivity to temperature variations compared to the conventional W-A configuration. This is particularly significant because, in the absence of anisotropy (\(\eta_m = 0\)), the noise suppression tends to degrade more significantly with temperature changes. By contrast, configurations with non-zero \(\eta_m\) maintain effective noise suppression across a much wider temperature range, making them more reliable for practical detection applications where temperature fluctuations are a concern. This finding underscores the potential advantages of anisotropic configurations in maintaining sensing accuracy and robustness in real-world scenarios where environmental conditions are rarely constant.
\subsection{Sensing Performance of the Probe Main Axis Along the x-direction}
Up to this point, our analysis has been limited to the case where the major axis of the ellipsoidal YIG sphere is aligned along the \(y\)-direction, yielding effective noise suppression within the optimal detection frequency range.  However, whether this favorable sensing performance persists is still unclear when the major axis is oriented along the \(x\)-direction. To address this question, we conduct a detailed investigation into the effects of this alternative configuration.

From Fig. \ref{Fig8}, it can be observed that when \(\eta_m\) is negative (i.e., the major axis of the ellipsoid is oriented along the \(y\)-direction), the system's performance undergoes significant changes. In Fig. \ref{Fig8}(a), the signal response \(A_m(\omega)\) gradually weakens as the absolute value of the negative parameter \(\eta_m\) increases. When \(\eta_m = 0\), the response amplitude is relatively high, maintaining a distinct peak near a normalized frequency of 1. However, as \(\eta_m\) becomes more negative, the overall amplitude of the response curve progressively decreases. Notably, when \(\eta_m = -0.9\omega_0\), the response strength is significantly reduced across the entire frequency range, indicating a diminished capability for signal amplification.
In Fig. \ref{Fig8}(b), the cavity field quantum noise \(N_{qn}(\omega)\) increases significantly with the increase in the negative value of \(\eta_m\). When \(\eta_m = 0\), the quantum noise remains relatively low, with only a slight rise near a normalized frequency of 1. However, for \(\eta_m = -0.5\omega_0\), \(-0.7\omega_0\), and \(-0.9\omega_0\), the overall magnitude of the noise curve increases markedly, with more pronounced peaks and fluctuations appearing over the normalized frequency range.
In Fig. \ref{Fig8}(c), the probe input thermal noise \(N_{mth}(\omega)\) also increases as the negative value of \(\eta_m\) increases. When \(\eta_m = 0\), the thermal noise is relatively low and increases gradually with frequency. However, as the absolute value of \(\eta_m\) increases, the growth rate of the thermal noise accelerates. Notably, the thermal noise level for \(\eta_m = -0.9\omega_0\) is significantly higher than for other parameter values.
In summary, when \(\eta_m\) is negative (i.e., when the major axis is aligned along the \(x\)-direction), the system’s signal response is significantly weakened. Meanwhile, the cavity field quantum and thermal noise increase with the absolute value of \(\eta_m\), indicating an overall deterioration in system performance.

Although the above analysis highlights that aligning the major axis along the \(x\)-axis is unfavorable for magnetic field sensing in the \(x\)-direction, a shift in perspective offers a new insight. For example, when the major axis is oriented along the \(y\)-axis, the system exhibits enhanced sensitivity to magnetic fields in the \(x\)-direction but reduced performance for detecting fields in the \(y\)-direction. Conversely, aligning the major axis along the \(x\)-axis improves sensitivity to \(y\)-direction magnetic fields while diminishing the ability to sense fields in the \(y\)-direction. This directional dependency reveals that the sensing precision can vary based on the relative alignment between the magnetic field and the ellipsoid's major axis, enabling directional sensing. Such a feature allows the system to selectively detect magnetic fields from a targeted direction while suppressing undesired signals from other directions. This anisotropic response could have significant practical applications, such as designing noise-resistant quantum sensors, creating directional magnetic field mapping systems, or enabling advanced signal filtering. By leveraging this directional sensing capability, the system can achieve enhanced performance and versatility, paving the way for innovative designs in magnetic sensing technologies.
\section{\label{sec4}Physical mechanism}
\begin{figure*}
	\centering\includegraphics[width=18cm,height=5cm]{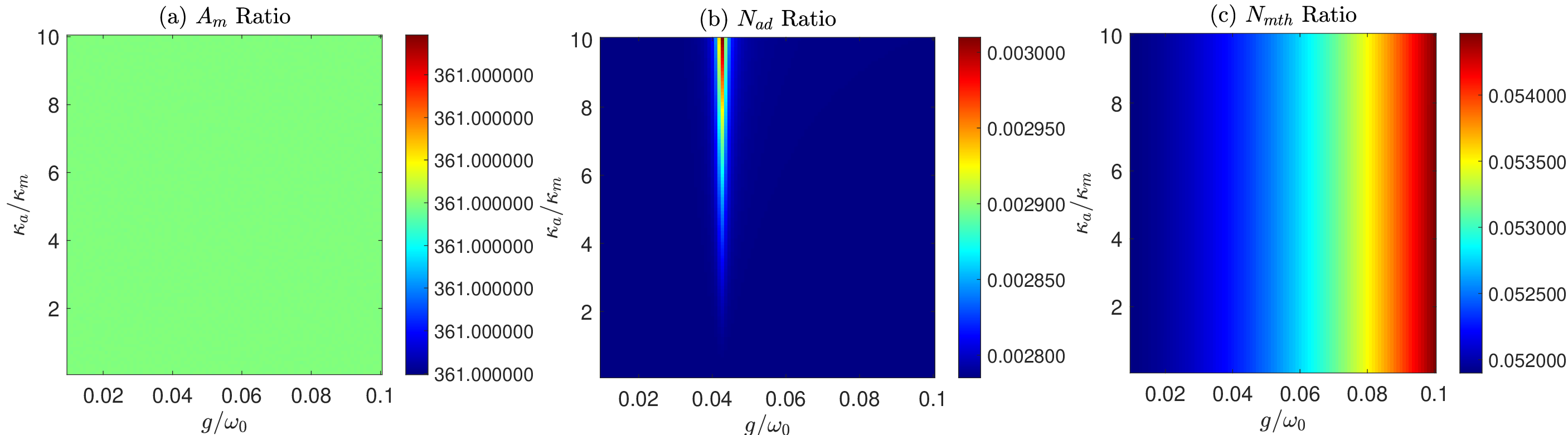}
	\caption{(a) The system's response ratio \(A_m(\omega)\) as a function of the normalized coupling strength \(g/\omega_0\) and the normalized dissipation rate \(\kappa_a / \kappa_m\). (b) The additional quantum noise ratio \(N_{ad}(\omega)\) plotted as a function of \(g/\omega_0\) and \(\kappa_a / \kappa_m\). (c) The thermal input noise ratio \(N_{mth}(\omega)\) as a function of \(g/\omega_0\) and \(\kappa_a / \kappa_m\). The initial environmental temperature is set at 5 mK, and the anisotropy parameter is fixed at \(\eta_m = \pm{0.9}\omega_0\).}
	\label{Fig9}
\end{figure*}

To comprehensively demonstrate the performance of directional sensing, we fixed \(\eta_m =\pm 0.9 \omega_0\) and analyze the ratios of three sensing performances along the \(x\)-axis and \(y\)-axis of the ellipse as functions of the normalized dissipation rate \(\kappa_a / \kappa_m\) and coupling strength \(g/\omega_0\). The results, presented in Fig. \ref{Fig8}, highlight several key features. In Fig. \ref{Fig9}(a), the system's response ratio between different directions reaches a value of 361, corresponding to two orders of magnitude (\(10^2\)), and remains remarkably robust across a wide range of \(\kappa_a / \kappa_m\) and \(g/\omega_0\). This demonstrates the stability and effectiveness of the directional amplification mechanism, which is largely insensitive to parameter variations. Fig. \ref{Fig9}(b) illustrates the ratio of the quantum noise contribution from the cavity field, which stays consistently low at approximately 0.0028 across the parameter space, with a slight enhancement near the optimal coupling strength of \(g \approx 0.04 \omega_0\). This highlights the system's strong isolation of cavity field quantum noise, ensuring minimal interference with the signal. In contrast, Fig. \ref{Fig9}(c) shows the ratio of the magnon thermal input noise from the probe input, which decreases at smaller cavity-magnon coupling strengths, indicating that higher directional sensing performance requires operation at lower coupling strengths to suppress magnon input thermal noise. Together, these results demonstrate the exceptional capability of directional sensing to achieve robust signal amplification and noise isolation over a broad range of parameters, making it a promising approach for high-precision sensing applications.

To elucidate the physical mechanism underlying the observed phenomenon and to account for the role of the anisotropic interaction of magnons in the system, we diagonalize the Hamiltonian \(H_1 = \hbar \omega_{0} \hat{m}^{\dagger} \hat{m} - \frac{\hbar \eta_{m}}{2}\left(\hat{m}^{2} + \hat{m}^{\dagger 2}\right)\) via the squeezing transformation \(\hat{M} = \cosh r_m \hat{m} + \sinh r_m \hat{m}^{\dagger}\), where the squeezing amplitude \(r_m = \frac{1}{4} \ln{\frac{\omega_0 - \eta_m}{\omega_{0} + \eta_m}}\) characterizes the anisotropy induced by \(\eta_m\). After the transformation, the Hamiltonian becomes
\begin{eqnarray}\label{13}
	\hat{H^{\prime}}&=&\hbar\Delta _a \hat{a}^{\dagger} \hat{a}+\hbar \omega^{\prime}_{0} \hat{M}^{\dagger} \hat{M}+\hbar g_1\left(\hat{a}^{\dagger}\hat{M}+\hat{a}\hat{M}^{\dagger}\right)\nonumber\\&+&\hbar g_2 \left(\hat{a}\hat{M}+\hat{a}^{\dagger}\hat{M}^{\dagger} \right)+i\hbar E_L( \hat{a}^{\dagger} -\hat{a})\nonumber\\&-&\hbar \lambda B^{\prime}_{ex}(t) (\hat{M}+\hat{M}^{\dagger}),
\end{eqnarray}
where \(\omega^{\prime}_0 = \omega_{0}/\cosh(2r_m)\) represents the renormalized effective magnon frequency, \(g_1 = g \cosh r_m\) and \(g_2 = -g \sinh r_m\) denote the modified cavity photon-magnon coupling strengths, and \(B^{\prime}_{ex}(t) = B_{ex}(t)e^{-r_m}\) represents the amplified effective external magnetic field. This squeezed representation provides a clear insight into the physical mechanisms at play. First, the renormalized magnon frequency \(\omega^{\prime}_0\) reflects the impact of anisotropy on the system’s dynamics and determines the location of key features in the response spectrum and cavity field quantum noise, such as peaks and dips. However, it is important to note that \(\omega^{\prime}_0\) is an approximate value, as the coupling strength \(g\) between the cavity photon and magnon modes introduces slight frequency shifts due to hybridization effects. These shifts result in a small deviation of the actual effective frequency from the theoretically predicted \(\omega^{\prime}_0\). 
Furthermore, after the transformation, \(g_1 = g \cosh r_m\) represents a beam-splitter-like interaction, while \(g_2 = -g \sinh r_m\) corresponds to a parametric amplification interaction. When the major axis of the ellipsoid is aligned along the \(y\)-axis, i.e., when \(\eta_m\) is positive, both interactions are enhanced due to the negative squeezing amplitude (\(r_m < 0\)), leading to a significant increase in the energy exchange between the cavity field and the magnon mode. This enhanced coupling is crucial in improving the system's ability to process weak signals, thereby boosting its overall sensitivity and noise suppression performance.
Additionally, through the squeezing transformation, the effective external magnetic field \(B^{\prime}_{ex}(t) = B_{ex}(t)e^{-r_m}\) is amplified. This amplification ensures that even weak external magnetic fields can induce a strong system's response, significantly improving the precision of magnetic field detection. The enhanced magnon mode amplified coupling strengths and the increased effective magnetic field form the foundation of the system's improved sensing capability.

Next, we demonstrate that the intrinsic anisotropy leads to a modification of the environment (reservoir) experienced by the squeezed magnon mode, which simultaneously introduces a corresponding noise suppression. This alteration results in a significant impact on the sensing performance. To further elaborate, we rigorously derive the quantum Langevin equation for the squeezed magnon mode, which can be given as
\begin{eqnarray}\label{14}
	\frac{d\hat{M}}{dt} &=& -i(\omega^{\prime}_0-i \frac{\kappa_m}{2}) \hat{M}(t) 
	+ \hat{F}_M(t).
\end{eqnarray}
For detailed information, please refer to Appendix~\ref{appendix:B}. 
Here, $\kappa_m$ denotes the dissipation rate of the YIG sphere magnon, and $\hat{F}_M(t)$ represents the input noise operator of the squeezed magnon, respectively. The correlation functions for the input operators are given as
\begin{align}
	\langle \hat{F}_M(t)\hat{F}_M(t') \rangle &=\kappa_m \sinh(2r_m)\left(\bar n_M + \frac{1}{2}\right)\delta(t-t'),\notag\\
	\langle \hat{F}_M(t)\hat{F}_M^\dagger(t') \rangle &= \kappa_m \delta(t-t') [\cosh(2r_m) \bar n_M+ \sinh^2(r_m)+1].
\end{align}
where $\bar n_a(\bar n_M)=[\exp(\hbar \omega _ { a }(\omega^{\prime} _ { 0 }) / k _ { B } T) - 1]^{-1}$ represents the average number of photons (squeezed magnons) in a thermal equilibrium state.
Using the method for solving noise in the system described earlier, we can derive the analytical expression for the total normalized output cavity field in the squeezed magnon mode representation as follows
\begin{widetext}
\begin{eqnarray}\label{19}
		\delta\hat{\mathcal{P}}^{\rm{out}}_{Na}(\omega)&=&\mathcal{A}_1(\omega)\hat{X}^{ \rm{in}}_a(\omega)+\mathcal{A}_2(\omega)\hat{P}^{ \rm{in}}_a(\omega)+\mathcal{A}_3(\omega)\hat{\tilde{X}}_{M}^{\rm in}(\omega)+\mathcal{A}_4(\omega)\hat{\tilde{P}}_{M}^{\rm { in }}(\omega)+\tilde{B}_{ex}(\omega),
\end{eqnarray}
where
\begin{eqnarray}
	\mathcal{A}_1 &=&  \frac{ \sqrt{\kappa_a} \left( 4\omega^2 \Delta_a + 4\omega^{\prime}_0 \left( (g_1 + g_2)^2 - \omega^{\prime}_0 \Delta_a \right) + 4i \omega \Delta_a \kappa_m - \Delta_a \kappa_m^2 \right) }{ 2 \sqrt{\kappa_m} \left( g_2 \left( -2i\omega(\omega^{\prime}_0 - \Delta_a) \right) + \omega^{\prime}_0 \kappa_a - \Delta_a \kappa_m \right) + g_1 \left( -2i\omega( \omega^{\prime}_0+ \Delta_a) \right) + \omega^{\prime}_0 \kappa_a + \Delta_a \kappa_m }, \notag\\
	\mathcal{A}_2 &=& \frac{
		 \Big( 16g_1^4 + 16g_2^4 + \big( 4\omega^2 - 4\Delta_a^2 + \kappa_a^2 \big) \big( -4\omega^{\prime 2}_0 + (2\omega + i\kappa_m)^2 \big) + 16g_2^2 \big( 2\omega^2 - 2\omega^{\prime}_0 \Delta_a + i\omega \kappa_m \big) - 16g_1^2 \big( 2(g_2^2 + \omega^2 + \omega^{\prime}_0 \Delta_a) + i\omega \kappa_m \big) \Big)
	}{
		8 \sqrt{\kappa_a} \sqrt{\kappa_m} \Big( g_2 \big( -2i\omega(\omega^{\prime}_0 - \Delta_a) + \omega^{\prime}_0 \kappa_a - \Delta_a \kappa_m \big) + g_1 \big( -2i\omega(\omega^{\prime}_0 + \Delta_a) + \omega^{\prime}_0 \kappa_a + \Delta_a \kappa_m \big) \Big)
	}, \notag\\
	\mathcal{A}_3 &=&\frac{
		e^{r_m} \Big( 
		4 (g_1 + g_2) \big( g_1^2 - g_2^2 - \omega^2 \big) 
		+ 4 (-g_1 + g_2) \omega^{\prime}_0 \Delta_a 
		- 2i (g_1 + g_2) \omega \kappa_a 
		+ (g_1 + g_2) (-2i \omega + \kappa_a) \kappa_m
		\Big)
	}{
		2 \Big(
		g_2 \big( -2i \omega (\omega^{\prime}_0 - \Delta_a) + \omega^{\prime}_0 \kappa_a - \Delta_a \kappa_m \big)
		+ g_1 \big( -2i \omega (\omega^{\prime}_0 + \Delta_a) + \omega^{\prime}_0 \kappa_a + \Delta_a \kappa_m \big)
		\Big)
	}, \notag\\
	\mathcal{A}_4 &=& e^{r_m},
\end{eqnarray}
and the output power spectrum density is given as
\begin{eqnarray}\label{21}
	S_{{\rm{out}}}(\omega)=N^{\prime}_{\rm{qn}}(\omega)+N^{\prime}_{mth}(\omega)+S_{\tilde{B}_{ex}}(\omega),
\end{eqnarray}
where $N^{\prime}_{\rm{qn}}(\omega) $, and $ N^{\prime}_{mth}(\omega) $ represent additional quantum noise of the cavity field, magnon input noise, respectively. Note that since the response is a common factor for both noise and signal, the response here has been normalized to 1, which can be achieved through an inverse filter. These noise are explicitly given as
\begin{align}\label{22}
	&N^{\prime}_{\rm{qn}}(\omega)=(\bar{n}_a+\frac{1}{2})e^{2r_m}( \vert {\mathcal{A}_1}(\omega)\vert ^2+ \vert {\mathcal{A}_2}(\omega)\vert ^2),\notag\\
	&N^{\prime}_{mth}(\omega)=(\bar{n}_M+\frac{1}{2})[e^{2 r_m}\vert {\mathcal{A}_3}(\omega)\vert ^2+1].
\end{align}
\end{widetext}
From the two noise expressions in Eq. (\ref{22}), it can be observed that the parameter \(r_m\) plays a critical role in controlling the noise suppression and signal amplification characteristics of the system. Directional sensing can be achieved by leveraging the relationship between the geometric spatial distribution of the YIG (yttrium iron garnet) sphere and compression. When \(r_m < 0\), the reduction of the factors \(e^{r_m}\) and \(e^{2r_m}\) significantly enhances the suppression of both quantum and thermal noise. This increases noise isolation along the \(y\)-axis. Conversely, when \(r_m > 0\), signals and noise are amplified to some extent. Due to the increase in \(e^{r_m}\) and \(e^{2r_m}\), the system's response along the \(x\)-axis is significantly enhanced, but at the cost of increased noise contributions, particularly along the \(y\)-axis. This interplay between noise isolation and signal amplification establishes the system's asymmetric behavior along the two axes of the ellipsoid, giving it a highly directional nature. From another perspective, if the magnetic field signal is distributed along both the \(x\)-axis and \(y\)-axis, based on the structure shown in Fig. \ref{Fig1}, weak magnetic field signals and noises along the \(x\)-axis will be improved, while signals and noise along the \(y\)-axis are unfavorable. Therefore, we achieve directional sensing, offering a robust solution under varying environmental conditions.
\label{sec5}\section{signal-to-noise ratio (SNR) and the magnetic field sensitivity}
\begin{figure}
	\centering\includegraphics[width=8.8cm,height=7cm]{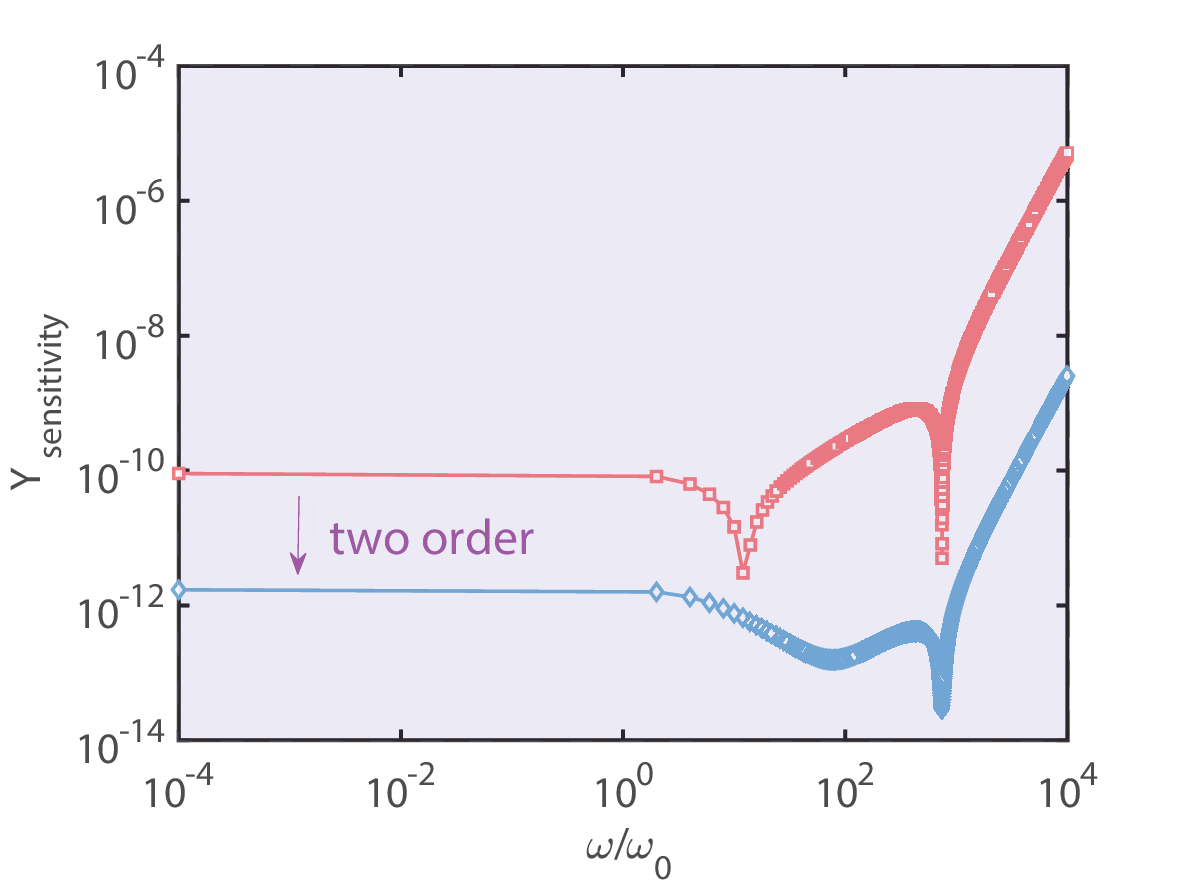}
	\caption{The system's sensitivity $Y_{sensitivity}$ as a function of the normalized frequency \(\omega/\omega_0\) is compared for squeezing parameters ($r_m$) of \(1.9\) and \(-1.9\), highlighting the advantage of directional sensing, with other parameters the same as those in Fig. \ref{Fig4}.}
	\label{Fig10}
\end{figure}

Finally, we discuss the signal-to-noise ratio and sensitivity of this system to better illustrate its overall performance.	To perform a detailed analysis of the sensitivity of this weak magnetic field sensing system, we consider the total noise spectral density after recovering the dimensionality. According to Eq. (\ref{19}), the total noise amplitude can be expressed as
\begin{equation}\label{23}
	\hat{B}_{\text{noise}}(\omega) = \frac{\kappa_{m}}{\sqrt{2}\lambda} \delta \hat{\mathcal{P}}^{\rm{out}}_{\text{Na}}(\omega) \, \Big|_{\tilde{B}_{\text{ex}}(\omega) = 0}
\end{equation}
The total noise intensity can be quantified using the symmetric noise power spectral density. This measure is directly detectable by a quantum spectrum analyzer and is defined as follows \cite{PhysRevA.89.053836}
\begin{equation}\label{24}
	S_{Bnoise}(\omega )\delta (\omega +\omega ^{\prime })=\frac{1}{2}(\langle \hat{
		B}_{noise}(\omega ) \hat{B}_{noise}(\omega ^{\prime })\rangle +c.c.).
\end{equation}
Thus, the total noise spectrum  can be given as
\begin{eqnarray}\label{25}
	S_{Bnoise}(\omega)=\frac{\kappa_m^2}{2\lambda^2}[N^{\prime}_{\rm{qn}}(\omega)+N^{\prime}_{mth}(\omega)].
\end{eqnarray}
To gain a deeper insight into the relationship between sensitivity and the effect of additional noise, we directly turn to the Signal-to-Noise Ratio (SNR) of the system defined by \cite{PhysRevA.103.062605, PhysRevA.100.023815}
\begin{eqnarray}\label{26}
\eta_{SNR}(\omega) = \frac{|B_S(\omega)|}{\sqrt{S_{Bnoise}(\omega)}}
\end{eqnarray}
where \(B_S(\omega)\) represents the total signal output of the system responding to the external magnetic field, and \(S_{Bnoise}(\omega)\) represents the total noise spectral density of the system. Note that \(\eta_{SNR} = 1\) means the response of the system to the external magnetic field is exactly the sensitivity \(Y_{\text{sensitivity}}(\omega)\), which represents the minimum detectable magnetic signal, equivalent to the scale of the ruler. Thus, it can be given as, with units of \( \text{T}/\sqrt{\text{Hz}} \).
\begin{eqnarray}\label{27}
	Y_{sensitivity}(\omega)&=&\sqrt{S_{Bnoise}(\omega)}\notag\\&=&\frac{\kappa_m\sqrt{[N^{\prime}_{\rm{qn}}(\omega)+N^{\prime}_{mth}(\omega)]}}{\sqrt{2}\lambda}.
\end{eqnarray}
 It is obvious that the sensitivity of the system depends on the additional cavity field noise and magnon  input thermal noise, the dissipation rate ($\kappa_{m}$) of the YIG sphere, and the coupling strength ($\lambda$) of the external magnetic field. A smaller $Y_{\text{sensitivity}}(\omega)$ value indicates higher sensitivity. To intuitively demonstrate the minimum detectable magnetic field and highlight the directional sensing capability, we present the sensitivity for squeezing parameters of \(\pm1.9\). As shown in Fig. \ref{Fig10}, the sensitivity with a squeezing parameter of \(-1.9\) (blue diamond markers) is nearly two orders of magnitude better than that with \(+1.9\) (red square markers), indicating a significant enhancement.This improvement allows our system to detect magnetic fields as low as $1~\mathrm{pT}$ near the magnon resonance frequency, and even down to $100~\mathrm{fT}$ in certain frequency ranges above the resonance, clearly showcasing the sensing advantage.

\section{\label{sec6} DISCUSSION and CONCLUSIONS}
In this paper, we propose a high-precision weak magnetic field sensing scheme based on geometric modulation and successfully realize magnon squeezing interactions using anisotropic yttrium iron garnet (YIG) spheres. Our study primarily addresses the challenges of signal amplification and noise suppression in directional weak magnetic field sensing within anisotropic cavity magnonic systems, particularly thermal noise from the input probe. To address this issue, we propose reshaping the YIG sphere from its conventional spherical geometry into an ellipsoidal shape, thereby inducing magnon squeezing. This geometric modification effectively adjusts the magnon mode's effective frequency, providing a parametric amplification effect that enhances the signal response while suppressing additional quantum noise from the cavity field and magnon input thermal noise. We also analyze the impact of the orientation of the ellipsoid's major axis on weak magnetic field sensing performance and propose the potential for directional sensing, which helps suppress undesired directional signals. This method significantly advances weak magnetic field sensing, offering a new pathway for substantially improving directional sensitivity.

In addition to its application in weak magnetic field sensing, cavity magnonic systems exhibit tremendous potential for dark matter detection. Specifically, interactions between axions and Standard Model particles can generate a pseudo-magnetic field, which magnons can detect. The resulting signal can be observed via optical readout techniques, as proposed in \cite{BARBIERI1989357}. Recent advancements in applying cavity magnonic systems for weak magnetic field sensing \cite{PhysRevLett.124.171801, FLOWER2019100306, Ruoso_2016, BARBIERI2017135} provide a robust platform for exploring their potential in dark matter detection. Future research will focus on leveraging quantum resources, such as entanglement, further to enhance the sensitivity of weak magnetic field sensing systems. These advancements can revolutionize weak magnetic field detection technologies, opening new frontiers in quantum sensing and fundamental physics research.
	\section*{Acknowledgments}
	We thank Chengsong Zhao, Ye-ting Yan and Shiwen He for their helpful discussions. This work was supported by the National Natural Science Foundation of China under Grant No. 12175029. 
		\section*{DATA AVAILABILITY}
	The data that support the findings of this article are not publicly available. The data are available from the authors upon reasonable request.
\appendix
\section{ANISOTROPICALLY SHAPED YIG SPHERE}\label{appendix:A}
This section evaluates \( H_F \) by quantizing the classical magnetic Hamiltonian \( H \). The Hamiltonian incorporates contributions from various physical interactions, including the Zeeman term $(H_Z)$, magnetocrystalline anisotropy $(H_{\text{aniso}})$, exchange interaction  ($H_{\text{ex}}$), and dipolar interaction ($H_{\text{dip}}$) as
\begin{eqnarray}\label{A1}
H = \int_V d^3r \, (H_Z + H_{\text{aniso}} + H_{\text{ex}} + H_{\text{dip}}),
\end{eqnarray}
where \( V \) is the volume of the ferromagnet \cite{peierls2001quantum,akhiezer1968spin}. To simplify the analysis, a sufficiently strong magnetic field \( B_0 \) is applied along the \( z \)-axis, ensuring the magnetic sphere reaches saturation magnetization. Additionally, when the sphere's dimensions are significantly larger than the domain wall length, the dipole-dipole interaction becomes the dominant factor in governing spin waves. As a result, the exchange interaction \( H_{\text{ex}} \) can be disregarded. For cubic materials, the influence of magnetocrystalline anisotropy is minimal and can also be neglected \cite{Prabhakar2009, Aharoni2001}.
The demagnetizing field of a uniformly magnetized ellipsoid can be written as
\begin{eqnarray}
	\mathbf{H}_{\text{dm}} = -(N_x M_x \hat{x} + N_y M_y \hat{y} + N_z M_z \hat{z}),
\end{eqnarray}
where \( N_{x,y,z} \) represents the demagnetization tensor and we have \( N_x + N_y + N_z = 1 \). The demagnetization tensor is related to the geometry of the ferromagnetic sphere. We set \( N_{y,z} = N \) and \( N_x = 1 - 2N \).
The dipolar interaction \( H_{\text{dip}} \) can be written as \cite{akhiezer1968spin}
\begin{eqnarray}
	H_{\text{dip}} = -\frac{\mu_0}{2} \mathbf{H}_{\text{dm}} \cdot \mathbf{M} \, ,
\end{eqnarray}
	where \( \mathbf{M} \) is the magnetization of the anisotropic-shaped YIG sphere. The classical magnetic Hamiltonian is given as
\begin{eqnarray}
	H = \int_V d^3r \, \left(-\frac{\mu_0}{2} \mathbf{H}_{\text{dm}} \cdot \mathbf{M} - \mu_0 \mathbf{M} \cdot \mathbf{B}_0 \right).
\end{eqnarray}
	The classical magnetic Hamiltonian is quantized by defining the magnetization operator \(\widetilde{\mathbf{M}} = -|\gamma|\widetilde{\mathbf{S}}\), where \(\widetilde{\mathbf{S}}\) is the spin density operator. The magnetization can be represented in terms of bosonic excitations through the application of Holstein-Primakoff transformations \cite{PhysRev.58.1098} as
\begin{eqnarray}
	\widetilde{M}_+ &=& \sqrt{2|\gamma|\hbar M_s}[1 - (|\gamma|\hbar/2M_s)\widetilde{m}_r^\dagger \widetilde{m}_r]\widetilde{m}_r,\notag\\
\widetilde{M}_- &=& \sqrt{2|\gamma|\hbar M_s}\widetilde{m}_r^\dagger [1 - (|\gamma|\hbar/2M_s)\widetilde{m}_r^\dagger \widetilde{m}_r],\notag\\
\widetilde{M}_z &=& M_s - |\gamma|\hbar \widetilde{m}_r^\dagger \widetilde{m}_r,
\end{eqnarray}
	where \(M_s\) is the saturation magnetization and \(\widetilde{M}_\pm = \widetilde{M}_x \pm i (|\gamma|/|\gamma|) \widetilde{M}_y\). The operator \(\widetilde{m}^{\dagger}_r = \sum_k \phi_k^*(r)b_k^\dagger\) is defined to create a magnon at position \(r\) with plane wave eigenstates \(\phi_k(r) = (1/\sqrt{V}) \exp(i \mathbf{k} \cdot \mathbf{r})\) and satisfies the bosonic commutation relation \([\widetilde{m}_r, \widetilde{m}_r^\dagger] = \delta(r - r')\).
therefore, the Hamiltonian \( \hat{H}_F \) can be expressed as
\begin{eqnarray}
	\hat{H}_F &= \left( \frac{\mu_0 |\gamma| M_s}{2} (N_x + N_y) - \mu_0 |\gamma| M_s N_z + \mu_0 |\gamma| B_b \right) \hat{m}^\dagger \hat{m}
\notag\\	&+ \frac{\mu_0 |\gamma| M_s}{4} (N_x - N_y)(\hat{m}^2 +\hat {m}^{\dagger 2}),
\end{eqnarray}
We focus on the \(\mathbf{k} = 0\) mode because it represents the uniform precession of magnetization and simplifies the Hamiltonian while capturing the dominant physical behavior, which can be given as
\begin{eqnarray}
	\hat{H}_F = \omega_0 \hat{m}^\dagger \hat{m} - \frac{\eta_m}{2} (\hat{m}^2 + \hat{m}^{\dagger 2}),
\end{eqnarray}
where \(\omega_0 = |\gamma| \mu_0 B_b + \eta_m\) and \(\eta_m = \frac{1}{2} \mu_0 |\gamma| M_s (1-3N)\). The sign of \(\eta_m\) is determined by the geometric shape of the ellipsoid, specifically by the distribution of the demagnetization factors \(N_x\), \(N_y\), and \(N_z\). When \(3N > 1\) (i.e., \(N_y = N_z = N > \frac{1}{3}\)), \(\eta_m < 0\), corresponding to an ellipsoid that is shorter along the \(y\)- and \(z\)-axes and longer along the \(x\)-axis. Conversely, when \(3N < 1\) (i.e., \(N_y = N_z = N < \frac{1}{3}\)), \(\eta_m > 0\), indicating an ellipsoid that is longer along the \(y\)- and \(z\)-axes and shorter along the \(x\)-axis. At the critical point where \(3N = 1\), \(\eta_m = 0\), corresponding to a spherical geometry with equal demagnetization factors in all directions, where the shape anisotropy vanishes, and the spin-wave frequency is solely determined by the external magnetic field \(B_0\). Thus, the sign and magnitude of \(\eta_m\) not only reflect the geometric properties of the ellipsoid but also govern its contribution to the spin-wave (magnon) frequency correction. In the main text, the correction to the magnon frequency caused by the demagnetizing factors can be compensated by the external magnetic field. Therefore, we assume it to be a constant.
	
To simplify and provide an explicit relationship between $\eta_m$ and the aspect ratio, we consider only a particular case. For other scenarios—including polynomial fits—please refer to Reference \cite{PhysRev.67.351,magnetism4030012}. a rotational spheroid (either a prolate or oblate spheroid), we assume its major axis is aligned along the \(x\)-direction, with rotational symmetry around the \(x\)-axis. The minor axes are along the \(y\)- and \(z\)-directions (\(y = z\)). The aspect ratio is defined as \(p =a/b\), where \(a\) is the semi-major axis and \(b\) is the semi-minor axis. The demagnetization factors \(N_x, N_y, N_z\) satisfy the symmetry condition \(N_x + 2N_y = 1\). We discuss two cases separately:
\section*{1. Prolate Spheroid (\(p > 1\))} The analytical expression for the demagnetization factor \(N_x\) is given by \cite{PhysRev.67.351}:
\begin{eqnarray}
	N_x &=& \frac{1}{p^2 - 1} \left[ \frac{p}{2\sqrt{p^2 - 1}} \ln\left( \frac{p + \sqrt{p^2 - 1}}{p- \sqrt{p^2 - 1}} \right) - 1 \right],\\
    N_y &=& N_z = \frac{1 - N_x}{2}.
\end{eqnarray}
Substituting into the expression for \(\eta_m\), defined as \(\eta_m = \frac{1}{2} \mu_0 |\gamma| M_s (1 - 3N_y)\), yields
\begin{eqnarray}
    \eta_m = \frac{3}{4} \mu_0 |\gamma| M_s \left( N_x - \frac{1}{3} \right).
\end{eqnarray}
 When \(p \to 1\) (sphere), \(N_x \to \frac{1}{3}\), and \(\eta_m \to 0\). When \(p \gg 1\) (needle-like spheroid), \(N_x \to 0\), and \(\eta_m \to -\frac{1}{4} \mu_0 |\gamma| M_s < 0\).
	\section*{2. Oblate Spheroid (\(p < 1\))}
	If the aspect ratio is less than 1, the analytical expression for the demagnetization factor \(N_x\) is given by \cite{PhysRev.67.351}
	\begin{eqnarray}
	N_x = \frac{1}{1 - p^2} \left[ 1 - \frac{p}{\sqrt{1 - p^2}} \arcsin\left( \sqrt{1 - p^2} \right) \right],
   \end{eqnarray}
	Substituting into the expression for \(\eta_m\) gives
	\begin{eqnarray}
	\eta_m = \frac{3}{4} \mu_0 |\gamma| M_s \left( N_x - \frac{1}{3} \right).
	\end{eqnarray}
When \(p \ll 1\) (disk-like spheroid), \(N_x \to 1\), and \(\eta_m \to \frac{3}{4} \mu_0 |\gamma| M_s \times \frac{2}{3} = \frac{1}{2} \mu_0 |\gamma| M_s > 0\). We can find that the sign and magnitude of \(\eta_m\) are directly controlled by the aspect ratio \(p\) via the demagnetization factor \(N_x(p)\). This relationship provides a theoretical foundation for experimental design, such as optimizing sensor sensitivity by selecting an appropriate aspect ratio.
\section{DERIVATION OF THE QUANTUM LANGEVIN EQUATION OF THE SQUEEZED MAGNON MODE}\label{appendix:B}
	In this section, we present a detailed, rigorous derivation of the quantum Heisenberg-Langevin equation of the squeezed magnon. the total Hamiltonian is $\hat{H} = \hat{H}_M + \hat{H}_B + \hat{H}_{M-B}$, as follows
	\begin{align}
		\hat{H}_M &=\hbar \omega^{\prime}_0 \hat{M}^\dagger \hat{M},  \\
		\hat{H}_B &=\hbar \int d\omega  \hat{B}^\dagger(\omega)\hat{B}(\omega), \\
		\hat{H}_{M-B} &= \hbar\int d\omega  g_{MB}(\omega)\left[ \hat{m}^\dagger\hat{B}(\omega)+\hat{m}\hat{B}^\dagger(\omega)  \right],
	\end{align}
where \(\hat{H}_M\) is the Hamiltonian of the squeezed magnon mode,
\(\hat{H}_B\) is the Hamiltonian of the bath,
\(\hat{H}_{M-B}\) is the interaction Hamiltonian between the original magnon mode and the environment thermal bath.
According to the squeezing transformation  $\hat{m} = \cosh r_m\hat{M}-\sinh r_m\hat{M}^{\dagger}$, the $\hat{H}_{M-B}$ can be given as
\begin{align}
	\hat{H}_{M-B} &= \hbar\int d\omega  g_{MB}(\omega)\left[ \hat{m}^\dagger\hat{B}(\omega)+\hat{m}\hat{B}^\dagger(\omega)  \right]\notag\\
	&=\hbar\int d\omega  g_{MB}(\omega)[ \hat{M}^\dagger\left(\cosh(r_m)B(\omega)-\sinh(r_m)\hat{B}^\dagger(\omega)\right)\notag\\&+\hat{M}\left(\cosh(r_m)\hat{B}^\dagger(\omega)-\sinh(r_m)\hat{B}(\omega)\right)].
\end{align}\label{B1}
\begin{widetext}
The Langevin equation is derived using the Heisenberg equations
\begin{align}
	\frac{d\hat{M}}{dt} &= -i\omega^{\prime}_0 \hat{M} -i \int d\omega g_{MB}(\omega) (	\cosh(r_m)\hat{B}(\omega) - \sinh(r_m)\hat{B}^\dagger(\omega)),\notag\\
	\frac{d\hat{B}(\omega)}{dt} &= -i\omega \hat{B}(\omega) -i g_{MB}(\omega) \left[\cosh(r_m)\hat{M} - \sinh(r_m)\hat{M}^\dagger\right].
\end{align}
Next, The formal solution to the \( \hat{B}(\omega)(t) \),\( \hat{B}^{\dagger}(\omega)(t) \) equation can be given as
\begin{align}
\hat{B}(\omega)(t)& = \hat{B}(\omega)(0)e^{-i\omega t} -i \int_0^t d\tau \, g_{MB}(\omega) [
\cosh(r_m)\hat{M}(\tau)- \sinh(r_m)\hat{M}^\dagger(\tau)] e^{-i\omega(t-\tau)},\notag\\
\hat{B}^\dagger(\omega)(t)& = \hat{B}^\dagger(\omega)(0)e^{i\omega t} +i \int_0^t d\tau \, g_{MB}(\omega) \big[
\cosh(r_m)\hat{M}^\dagger(\tau) - \sinh(r_m)\hat{M}(\tau)
\big] e^{i\omega(t-\tau)}.
\end{align}
Substitute the formal solution for $ \hat{B}(\omega) $, $\hat{B}^\dagger(\omega)$ into the equation for \( \hat{M} \), we get
		\begin{align}
			\frac{d\hat{M}}{dt} = & -i\omega^{\prime}_0 \hat{M} - i \int d\omega g_{MB}(\omega) \Bigg\{ 
			\cosh(r_m) \Bigg[\hat{B}(\omega)(0)e^{-i\omega t} - i \int_0^t d\tau g_{MB}(\omega) \big[\cosh(r_m)\hat{M}(\tau) \notag \\
			& - \sinh(r_m)\hat{M}^\dagger(\tau)\big]e^{-i\omega(t-\tau)} \Bigg]
			- \sinh(r_m) \Bigg[\hat{B}^\dagger(\omega)(0)e^{i\omega t} + i \int_0^t d\tau g_{MB}(\omega) \big[\cosh(r_m)\hat{M}^\dagger(\tau) \notag \\
			& - \sinh(r_m)\hat{M}(\tau)\big]e^{i\omega(t-\tau)} \Bigg] \Bigg\}.
		\end{align}
It can be rewritten as
\begin{align}
	\frac{d\hat{M}}{dt} = -i\omega^{\prime}_0 \hat{M} -i \int d\omega g_{MB}(\omega) \Big[
	\cosh(r_m)\hat{B}(\omega)(0)e^{-i\omega t} - \sinh(r_m)\hat{B}^\dagger(\omega)(0)e^{i\omega t}
	\Big]
	- \int_0^t d\tau \, K_1(t-\tau),
\end{align}
	where the memory kernel \( K_1(t-\tau) \) is:
\begin{align}
	K_1(t-\tau) &= \cosh^2(r_m)\int d\omega g_{MB}^2(\omega) e^{-i\omega(t-\tau)} \hat{M}(\tau) - \sinh^2(r_m)\int d\omega g_{MB}^2(\omega) e^{i\omega(t-\tau)} \hat{M}(\tau)\notag\\&-\cosh(r_m)\sinh(r_m)\int d\omega g_{MB}^2(\omega) e^{-i\omega(t-\tau)} \hat{M}^\dagger(\tau)\notag\\&+\cosh(r_m)\sinh(r_m)\int d\omega g_{MB}^2(\omega) e^{i\omega(t-\tau)} \hat{M}^\dagger(\tau).
\end{align}
	\end{widetext}
To simplify, let us define
\begin{align}
G^+(t-\tau) &= \int d\omega g_{MB}^2(\omega) e^{-i\omega(t-\tau)}, \notag\\
G^-(t-\tau) &= \int d\omega g_{MB}^2(\omega) e^{i\omega(t-\tau)}.
\end{align}
Using these expressions, the memory kernel can be rewritten as
\begin{align}
	K_1(t-\tau) &= \cosh^2(r_m) G^+(t-\tau) \hat{M}(\tau) - \sinh^2(r_m) G^-(t-\tau) \hat{M}(\tau) \notag\\
	& - \cosh(r_m)\sinh(r_m) G^+(t-\tau) \hat{M}^\dagger(\tau) \notag\\
	&+ \cosh(r_m)\sinh(r_m) G^-(t-\tau) \hat{M}^\dagger(\tau).
\end{align}
In most physical scenarios, the coupling strength squared, \( g_{MB}^2(\omega) \), is symmetric concerning frequency. As a result, we have  
\begin{align}
G^-(t-\tau) = \big[G^+(t-\tau)\big]^*.
\end{align}
Substituting this symmetry into the expression gives
\begin{align}
	K_1(t-\tau) &= \cosh^2(r_m) G^+(t-\tau) \hat{M}(\tau) - \sinh^2(r_m) \big[G^+(t-\tau)\big]^* \hat{M}(\tau) \notag\\
	& - \cosh(r_m)\sinh(r_m) G^+(t-\tau) \hat{M}^\dagger(\tau) \notag\\
	& + \cosh(r_m)\sinh(r_m) \big[G^+(t-\tau)\big]^* \hat{M}^\dagger(\tau).
\end{align}
 Expand \( G^+(t-\tau) \) and \( G^-(t-\tau) \) as
\begin{align}
G^+(t-\tau) = R(t-\tau) + iI(t-\tau),
\end{align}
where \[
R(t-\tau) = \operatorname{Re}[G^+(t-\tau)] \quad \text{and} \quad I(t-\tau) = \operatorname{Im}[G^+(t-\tau)],
\]  
represent the real and imaginary parts of \( G^+(t-\tau) \), respectively.
Similarly, the conjugate is
\begin{align}
\big[G^+(t-\tau)\big]^* = R(t-\tau) - iI(t-\tau).
\end{align}
Combining all terms, the simplified memory kernel becomes as
\begin{align}
	K_1(t-\tau) &= R(t-\tau) \big[\hat{M}(\tau) - 2i\cosh(r_m)\sinh(r_m) I(t-\tau)\hat{M}^\dagger(\tau)\big] \notag\\
	&\quad + iI(t-\tau)\big(\cosh^2(r_m) + \sinh^2(r_m)\big)\hat{M}(\tau),
\end{align}
For the real part \( R(t-\tau) = \operatorname{Re}[G^+(t-\tau)] \), under the Markov approximation, $R(t-\tau) \approx \kappa_m \delta(t-\tau)$.
For the imaginary part \( I(t-\tau) = \operatorname{Im}[G^+(t-\tau)] \), this term typically represents a frequency shift. If included, it can also be approximated as:
$I(t-\tau) \approx \Delta \delta(t-\tau),$
Where \( \Delta \) corresponds to an effective frequency shift from the imaginary part. Substituting the Markov approximation, the memory kernel becomes as
\begin{align}
	K_1(t-\tau) &= \kappa_m \delta(t-\tau) \hat{M}(\tau)\notag\\
	& - 2i \Delta\delta(t-\tau)\cosh(r_m)\sinh(r_m) \hat{M}^\dagger(\tau) \notag\\
	&\quad + i\Delta \delta(t-\tau)\big(\cosh^2(r_m) + \sinh^2(r_m)\big)\hat{M}(\tau).
\end{align}
The \( \delta(t-\tau) \) ensures that the memory kernel's influence is instantaneous, affecting only the current time. Then, substituting the simplified memory kernel into the original Langevin equation as
\begin{align}
\frac{d\hat{M}}{dt} = -i\omega^{\prime}_0 \hat{M} - \int_0^t d\tau K_1(t-\tau),
\end{align}
thus we yield
\begin{align}
\frac{d\hat{M}}{dt} &= -i\omega^{\prime}_0 \hat{M} - (\kappa_m/2) \hat{M}(t) - i\Delta\cosh(r_m)\sinh(r_m)\hat{M}^\dagger(t)
\notag\\&- i\Delta/2 \big(\cosh^2(r_m) + \sinh^2(r_m)\big)\hat{M}(t) + F_M(t).
\end{align}
The above can also be simplified as
\begin{align}
\frac{d\hat{M}}{dt} &= -i\big(\omega^{\prime}_0 + \frac{\Delta}{2}(2\cosh^2(r_m) - 1)\big)\hat{M}(t) 
- \frac{\kappa_m}{2} \hat{M}(t)\notag\\ 
&-i\Delta \cosh(r_m)\sinh(r_m)\hat{M}^\dagger(t) 
+ F_M(t),
\end{align}
where \( F_M(t) \) is the noise term, describing random fluctuations from the environment, which can give as
\begin{align}
	F_M(t) &= -i\int d\omega g_{MB}(\omega) [
	\cosh(r_m)\hat{B}(\omega)(0)e^{-i\omega t} \notag\\&- \sinh(r_m)\hat{B}^\dagger(\omega)(0)e^{i\omega t}].
\end{align}
Using the bath expectation value as
\begin{eqnarray}
\langle \hat{B}^\dagger(\omega)\hat{B}(\omega') \rangle &=& n(\omega)\delta(\omega-\omega'),\notag\\
\langle \hat{B}(\omega)\hat{B}^\dagger(\omega') \rangle &=& (n(\omega)+1)\delta(\omega-\omega'), 
\end{eqnarray}
and assuming the coupling strength \( g_{MB}^2(\omega) \) is slowly varying, and the thermal occupation \( n(\omega) \) can be approximated by its value \( \bar n_M \) at a central frequency ($\omega^{\prime}_{0}$), the integral simplifies to 
\begin{eqnarray}
\int d\omega g_{MB}^2(\omega)n(\omega)e^{i\omega(t+t')} \approx \kappa_m \bar n_M \delta(t+t'),
\end{eqnarray}
Substituting this back, the  correlation functions  can be given as
\begin{align}
\langle \hat{F}_M(t)\hat{F}_M(t') \rangle &=\kappa_m \sinh(2r_m)\left(\bar n_M + \frac{1}{2}\right)\delta(t-t'),\notag\\
\langle \hat{F}_M(t)\hat{F}_M^\dagger(t') \rangle &= \kappa_m \delta(t-t') [\cosh(2r_m) \bar n_M+ \sinh^2(r_m)+1].
\end{align}
In practical problems, the frequency shift effect can often be neglected, allowing us to ignore the \(\Delta\) term. The resulting Langevin equation is then given as
\begin{eqnarray}
		\frac{d\hat{M}}{dt} &=& -i(\omega^{\prime}_0-i \frac{\kappa_m}{2}) \hat{M}(t) 
		+ \hat{F}_M(t).
\end{eqnarray}
This equation characterizes the squeezed magnon mode dynamics of the system.
\begin{widetext}
Next, we introduce the all correlation function for the input noise, which is defined as follows according to Equation (B24)
\begin{align}\label{A2}
	&\left\langle \hat F_{M}(t) \hat F_{M}^{\dagger}\left(t^{\prime}\right)\right\rangle= \kappa_m [\cosh(2r_m)\bar n_M+\sinh^2(r_m)+1]\delta\left(t-t^{\prime}\right),  \notag\\
	&\left\langle \hat F_{M}^{\dagger}(t) \hat F_{M}^{\dagger}\left(t^{\prime}\right)\right\rangle= \kappa_m \sinh(2r_m)(\bar n_M+1/2)\delta\left(t-t^{\prime}\right),\notag\\
	&\left\langle \hat F_{M}(t) \hat F_{M}\left(t^{\prime}\right)\right\rangle= \kappa_m \sinh(2r_m)(\bar n_M+1/2)\delta\left(t-t^{\prime}\right),\notag\\
	&\left\langle \hat F_{M}^{\dagger}(t) \hat F_{M}\left(t^{\prime}\right)\right\rangle=\kappa_m [\cosh(2r_m)\bar n_M+\sinh^2(r_m)]\delta\left(t-t^{\prime}\right).
\end{align}
We also present the quadrature components of the amplitude and phase for the squeezed magnon mode, which can be expressed as follows
\begin{eqnarray}\label{A3}
	\hat{X}_{M}^{\rm in}&=&(\hat{M}_{in}^{\dagger }+\hat{M}_{in})/\sqrt{2},\notag\\
	\hat{P}_{M}^{ \rm in}&=&(\hat{M}_{in}-\hat{M}_{in}^{\dagger })/\sqrt{2}i.
\end{eqnarray}
The correlation functions for the amplitude and phase quadrature components of the squeezed magnon mode are given as follows
\begin{align}\label{A4}
	\left\langle \hat{X}_{M}^{\rm in}(t) \hat{X}_{M}^{\rm in}\left(t^{\prime}\right)\right\rangle&=\frac{1}{2}\left[\left\langle \hat F_{M}(t) \hat F_{\rm M}\left(t^{\prime}\right)\right\rangle+\left\langle \hat F_{M}(t) \hat{F}_{M}^{\dagger}\left(t^{\prime}\right)\right\rangle+\left\langle \hat F_{M}^{\dagger}(t) \hat F_{M}\left(t^{\prime}\right)\right\rangle+\left\langle \hat F_{M}^{\dagger}(t) \hat F_{M}^{\dagger}\left(t^{\prime}\right)\right\rangle\right]\notag\\ 
	&=\frac{\kappa_m}{2}\left[2\sinh(2r_m)(\bar n_{M }+1/2)+2\cosh(2r_m)\bar n_{M }+\cosh(2r_m)\right]
	\notag\\ 
	&=\frac{\kappa_m}{2}\left[2\sinh(2r_m)(\bar n_{M }+1/2)+2\cosh(2r_m)(\bar n_{M}+1/2)\right]	\notag\\ 
	&=\kappa_m e^{2r_m}(\bar n_{M }+1/2)\delta\left(t-t^{\prime}\right).
\end{align}
\begin{align}
	\left\langle \hat{P}_{M}^{\rm in}(t) \hat{P}_{M}^{\rm in}\left(t^{\prime}\right)\right\rangle&=-\frac{1}{2}\left[\left\langle \hat F_{M}(t) \hat F_{\rm M}\left(t^{\prime}\right)\right\rangle-\left\langle \hat F_{M}(t) \hat F_{M}^{\dagger}\left(t^{\prime}\right)\right\rangle-\left\langle \hat F_{M}^{\dagger}(t) \hat F_{M}\left(t^{\prime}\right)\right\rangle+\left\langle \hat F_{M}^{\dagger}(t) \hat F_{M}^{\dagger}\left(t^{\prime}\right)\right\rangle\right]
	\notag\\ 
	&=-\frac{\kappa_m}{2}\left[2\sinh(2r_m)(\bar n_{M }+1/2)-2\cosh(2r_m)\bar n_{M }-\cosh(2r_m)\right]\notag\\ 
	&=-\frac{\kappa_m}{2}\left[2\sinh(2r_m)(\bar n_{M }+1/2)-2\cosh(2r_m)(\bar n_{M }+1/2)\right]
	\notag\\ 
	&=\kappa_m e^{-2r_m}(\bar n_{M }+1/2)\delta\left(t-t^{\prime}\right).
\end{align}
\end{widetext}

\bibliography{magnetometry}
\end{document}